# Is dark matter in spiral galaxies cold gas?

## II. Fractal models and star non-formation

**Daniel Pfenniger[1], Françoise Combes[2]**

[1] Observatoire de Genève, CH-1290 Sauverny, Switzerland
[2] DEMIRM, Observatoire de Meudon, 92 190 Meudon, France

**Abstract.** In a companion paper (Paper I) we have proposed a new candidate to account for the dark matter around spiral galaxies: cold $H_2$ gas in a fractal structure, supported by rotation, and concomitant with the HI disc. We have shown that this hypothesis is compatible with dynamical and observational constraints about disc galaxies, and explains several conspiracies and paradoxes, since the dark matter is then in a form of fresh gas able to produce stars. In this paper we attempt to describe the physical conditions leading to a fractal state of cold gas in outer galaxy discs.

Gas cloud models taking into account the recently disclosed fractal structure of cold gas are set up, showing that large errors in the classical gas mass determination based on smooth cloud models can easily follow if the gas is in reality fractal. Indeed the range of possible column densities is then much larger, including 5 or more decades of surface densities, instead of 2 for smooth cloud models. Thus fractal clouds must present both optically thin and optically thick clumps in any single wavelength observations. The observed fractal dimension of the cold ISM suggests that mass underestimates by a factor 10 or more are typical. Due to its low temperature (around 3 K), and its condensed fractal structure, together with its low metallicity, the outer gas would be almost invisible for usual detectors.

We consider the paradox of the persistence of cold Jeans unstable gas in outer discs, far from important heating sources, yet not forming stars or Jupiters. Following Rees (1976), we determine the smallest clump distribution that can persist in a collisional and almost isothermal fragmenting cold gas. At 3 K these elementary cloudlets are predicted to have a radius of about 30 AU, and have a mass of the order of a Jupiter. Their average density and column density are $10^9 \, \mathrm{cm}^{-3}$ and $10^{24} \, \mathrm{cm}^{-2}$. They are gravitationally bound, and their line of sight thermal width is about $0.1 \, \mathrm{km \, s}^{-1}$. Their frequent collisions prevent them from forming Jupiters or stars and the near isothermality of the fractal nearly suppresses energy dissipation. At higher temperature, especially above $H_2$ dissociation, the collision rate in the fractal decreases, favouring star formation.

It turns out that the smallest density condensations, called "clumpuscules" offer favourable conditions for containing $H_2$ in both vapour and solid phases. However it is unknown whether enough condensations sites such as dust exist in the outer discs to permit the freezing of $H_2$. It is expected that the large sublimation energy prevents much $H_2$ to become solid, but a small amount of $H_2$ ice grains is a crucial factor for a good coupling between gas and the 3 K background.

Many of the general arguments presented here about fractals can be applied to other inhomogeneous structures, such as the hot gas in galaxy clusters. The clumpuscules presented here might be the form of matter in which cooling flows in clusters seem to disappear.

**Key words:** dark matter – galaxies:ISM – ISM: structure – stars: formation

## 1. Introduction

In order to understand better the problems and conspiracies discussed in Paper I (Pfenniger et al. 1994), we consider in more depth its main hypothesis: dark matter in spiral galaxies would be made essentially of a form of hydrogen sufficiently diffuse to be still able to form stars, and at the same place than HI, mainly outside the optical discs. It should be preferentially in cold molecular gas form, since the warm and hot diffuse phases fill most of the volume at a too low density to contribute significantly to dark matter. Indeed the existence of a cold, dense, self-shielded and very low volume-filling gas is compatible with observations. This is the only gaseous form where a large amount of mass can be hidden. In the numerous conferences dedicated to the dark matter problem, extremely little critical attention has been devoted to this possibility. This paper contains in more detail a work presented elsewhere as a first account (Pfenniger 1993).

First of all, the fact that hydrogen at low temperature has very short cooling times ($\ll 10^6$ yr), is Jeans unstable, but manages to survive in the outer discs of galaxies for several Gyr remote from important heating sources and without collapsing into stars or Jupiters, is a fact not properly understood. Apparently, cold and clumpy gas with supersonic turbulence velocities ($\approx 10$ km s$^{-1}$) dissipates very little. This problem has been mentioned several times in the context of molecular clouds in the optical part of galaxies (see e.g. Scalo 1985), but since stars are then possible important energy input sources, the long molecular cloud lifetime there is not a problem as obvious as for HI in outer galactic discs.

The physical state of this gas must be high density and low temperature. Since no significant heating sources in the outer galactic discs presumably exist, we can assume that the gas is only bathing in the cosmological background, and that its temperature is about 3 K. In these nearly isothermal conditions, clouds can fragment (Hoyle 1953) until they reach small clump units, in which the cooling time becomes comparable to the free-fall time (Rees 1976). The average typical density of these elementary cloudlets, called "clumpuscules", is $10^9$ cm$^{-3}$, column density $10^{24}$ cm$^{-2}$, size 30 AU, and mass $10^{-3}$ M$_\odot$. These small units are the building blocks of a fractal structure, that ranges upwards over 4 to 6 orders of magnitude in scales. They are gravitationally bound, and their individual thermal width for molecular hydrogen along the line of sight is about 0.1 km s$^{-1}$.

Based on the fact that cold gas is observed to be fractal over several decades of length and density, models of fractal clouds are built by Monte-Carlo and their projection properties are examined, leading to understand the origin of possible large underestimates in the cold gas mass. As a first step, simple static hierarchical mass distributions à la von Hoerner (cf. von Weizsäcker 1951; Hoyle 1953) are built by Monte-Carlo simulations. Despite their simplicity, the models here mimic the inhomogeneity and self-similar properties of real clouds much better than homogeneous classical cloud models. Clearly real clouds might be more complex, e.g. multifractal (Chappell & Scalo 1994), but simple models are useful to grasp the first order effects introduced by *essentially* non-smooth mass distributions.

For the sake of clarity, we have deliberately not mentioned many additional physical complications, such as magnetic fields, that inevitably exist in reality. These should certainly be taken into account in further investigations.

## 2. Fractal geometry of cold gas

### 2.1. Cold ISM fractal structure

In recent years the fractal nature of the interstellar cold gas has been increasingly well documented (Falgarone et al. 1992). Nearby molecular clouds are observed self-similar in projection over a range of scales and densities of at least $10^4$ (Scalo 1985; Falgarone et al. 1991; Elmegreen 1992), but perhaps up to $10^6$. Indeed, VLBI Galactic HI absorption measures show structures on a scale as low as 25 AU (Diamond et al. 1989), and quasar monitoring reveals nearby $\approx 10$ AU sized clouds scattering strongly radio-waves (Fiedler et al. 1987). Moreover, very high densities, up to about $10^9$ cm$^{-3}$, are observed near star-forming regions, in solar-system sized condensations (e.g. Churchwell et al. 1987; Stutzki et al. 1988; Wilking et al. 1989; Mundy et al. 1992). Other spectacular examples of small and dense condensations are the cometary globules in the Helix planetary nebula (Walsh & Meaburn 1993) that seem to be evaporated by the central hot star. Very recently Marscher et al. (1993) have claimed to detect AU-scale clumps at high density ($10^7$ cm$^{-3}$) in Galactic molecular clouds by monitoring extragalactic sources.

Therefore the smallest detected structures have some tens of AU, the largest structures still only weakly perturbed by galactic rotation, a few 100 pc. Remarkably the fractal dimension of *isophotes* at different wavelengths is almost constant, around $1.3 - 1.4$, over a wide range of scales, and independent of the molecular or atomic state of the gas (Bazell & Désert 1988; Scalo 1985, 1990; Falgarone 1992).

### 2.2. Fractal geometry

In nature a large variety of fractal-like objects with similar fractal dimensions are known, that can be driven by turbulence, chaos, self-gravity, percolation, or chemical processes. So the fractal structure of the ISM in itself cannot inform us unequivocally about the predominant physical mechanism causing it. (In classical geometry too, similar forms, e.g. spheres, can arise from widely different physical reasons.)

The projection properties of fractals are insufficiently studied. The *section* of a fractal of dimension $D$ is known to have generally a dimension $D_s = D - 1$. However *projection* is a different operation that follows another rule. For example the section of a line ($D = 1$) is typically a point ($D_s = 0$) but the projection of a line is typically a line ($D_p = 1$). The section and projection of smooth density distributions ($D = 3$) have generally $D_s = D_p = 2$. For fractal distributions with $D < 3$ an intermediate behaviour is expected.

An accurate account of the existing mathematics about fractal projection is given by Falconer (1985, 1990 Chap. 6). For a fractal set of dimension $D$ in space and for a projection onto a plane we have that typically

$$D_p = \begin{cases} D & \text{for } D \leq 2, \\ 2 & \text{for } D \geq 2. \end{cases} \tag{1}$$

This is different from the formula $D_p = D_s = D - 1$ sometimes used.

It should be noted that, contrary to a common belief, a mass distribution with an integer dimension lower than 3 does not imply that the distribution is necessarily similar to a set of surfaces, or of lines[1]. This is related to the fact that the fractal (Hausdorff) dimension just describes how the *number* of mass elements varies with distance, it does not constraint neighbouring points to be smoothly distributed over a surface or along a line. Beside $D$, many other parameters can characterise a mass distribution.

---

[1]   Mandelbrot's book (1982) contains examples of such integer dimensioned but fractal objects: Fournier Universe, $D = 1$ p. 96, and a skewed web, $D = 2$ p. 142.

### 2.3. Cold ISM fractal dimension

Thus the direct application of mathematics to observations of cold gas leads to infer that in the ISM $D \approx 1.4$. But, contrary to simple fractal models, either opaque or transparent, interstellar gas presents a distribution of optical depths, which may modify the relation between the apparent and the real dimension.

In any case the general fractal isophotes of nearby molecular clouds over a large range of scales and physical properties strongly suggest that cold HI in the outer discs may also be better described by a fractal model than by "standard" smooth clouds. Virial equilibrium arguments applied to hierarchical clouds yield an explanation to the observed Larson's relations: the mass-radius relation $M \sim r^D$, and the velocity dispersion-radius relation $\sigma^2 \sim r^{D-1}$ (Larson 1981, 1992; Myers 1985; Scalo 1985; Chièze 1987).

From recent observational data these relations indicate a fractal dimension $D \sim 1.6 - 2$. Solomon et al. (1987) propose that the mass of a molecular cloud varies as the square of its radius, suggesting $D = 2$. However, this is not a relation deduced directly from observations. The main observation is the size line-width relation, i.e. $\sigma$ varying as about $r^{0.5}$. Other relations have been observed, like $\sigma$ varying as $r^{0.3}$ (e.g. Larson 1981; Falgarone & Phillips 1992), revealing that the way clouds sizes are defined can change significantly the resulting power-law (e.g. Combes 1991). With an exponent of 0.3, the deduced virial mass, varying as $r^{1.6}$, would suggest $D = 1.6$. Scalo (1987) argued that $D$ should be less than 2 because $D = 2$ would imply a constant column density, while smaller clouds are observed to be more opaque.

In fact we don't have any compelling theoretical reason for a universal $D$. On the contrary a continuity argument suggests $D$ to be variable in time: when a diffuse cloud starting with $D = 3$ begins to fragment, it doesn't reach immediately a fractal state with a constant $D$ everywhere. If star formation occurs and stirs the medium by a large energy input, the resulting structure is likely to be different from the one that is produced in cold regions where no detectable star formation happens.

### 2.4. Modelling fractal structures

Modelling the physics of fractal objects seems presently far from trivial. Indeed up to recently physics has focused on investigating smooth objects. As classical geometry, fractal geometry is an idealisation of the shape of physical systems that breaks down at some small scale. Unless we have the instrumental or the computing power to resolve a fractal down to the smallest scale, below which a smooth behaviour might be applicable, one of the most basic traditional assumption in physics, differentiability, is lost. So all the traditional tools of calculus such as differential equations, cannot be used as directly as in smooth systems. At the present time mathematical tools allowing to work on the physics of fractal objects are rudimentary (see Falconer 1990 for a good overview).

Often some physical properties greatly differ between a smooth and fractal objects. For example (e.g. Wright 1993) an interstellar grain can cool much faster if its "surface" is as rough as a fractal, because the cooling time depends on the surface size. In a strict mathematical fractal grain model the surface would be infinite, in a real grain the effective surface depends on the smallest scale at which the fractal behaviour stops.

So if the cold ISM is fractal, unless the smallest scale is accessible to observations (respectively for models, to calculations), where differentiability might be a valid assumption, differential equations may not be applied. Even if the ISM is made mostly of gas in which the molecule mean-free path is always much shorter than a pc, the use of the hydrodynamic or radiation transfer equations is justified only when the resolution in space and velocity is high enough to treat or detect the smallest clumps.

Although the general evidence is that HI is structured down to the smallest accessible scales (see e.g. Burton 1992), lacking better tools observers apply the equation of transfer and assume homogeneity at subresolution scales to estimate the amount of interstellar gas, which in turn provides the amount of baryons in the outer discs of spirals.

### 2.5. Gas modelling in galaxies

Modelling "correctly" the gas dynamics at the scale of galaxies has been recognised to be a non-trivial problem many times (e.g. Prendergast 1962, Binney & Gerhard 1993, Sellwood & Wilkinson 1993). This is not astonishing if the gas is fractal because then the smallest scale of the fractal requires an inaccessible high resolution (Scalo 1985), and as consequence the only reason to pick out a particular smallest scale is the computer capacity.

The other problem of modelling the cold ISM gas is the general turbulence (the Reynolds number is large) associated with supersonic motion (the Mach number is large). In a generally smooth fluid, a shock is an exceptional region where the differentiability condition fails, requiring a separate treatment by jump conditions. When a fluid is supersonically turbulent, one expects generalised shocks, the differentiability of the flow is violated almost everywhere. The smooth hydrodynamical model of fluid flow is generally not applicable, but no other conceptual tool seems today to exist.

In extreme inhomogeneous fluid systems, say modelling the motion of Jupiter through the interplanetary plasma, or the motion of stars through the ISM, it is clear that a global hydrodynamical description is not necessary; a distinct treatment of the two extreme regimes is good modelling. In less extreme but still very inhomogeneous gaseous systems such as the cold ISM, the distinction is less obvious, but still if the density contrast between the densest and the most tenuous lumps of gas spans several decades over a few AU (as claimed recently by Marscher et al. 1993), clearly an hydrodynamical approach is a priori difficult. Quoting Scalo (1985): "... once we admit that the interstellar medium is turbulent, there is very little theoretical basis on which to proceed". Clearly new kinds of models are required for the cold ISM. A first order attempt stressing the self-similar aspect of the ISM instead of the hydrodynamic one, aimed at representing the extraordinary inhomogeneity of the ISM, is given in the next Section.

At the simplest level the galactic gas has often been modelled as a smooth and finite distribution of atoms $\rho(\boldsymbol{x})$, this is a one-level hierarchical structure: a system is decomposed into a large number of identical sub-systems. In principle if the equation of state is known, hydrodynamical equations can be used. Recognising that the ISM has sub-structures, further refinements introduced "standard clouds" (e.g. Spitzer 1956), the galactic gas is then a distribution of spherical clouds, each one of them being a smooth distribution of atoms, a two-level hierarchy.

Typically the mean-free path of clumps at the $10 - 50\,\mathrm{pc}$ scale, such as molecular clouds, is much larger than their size, an hydrodynamical description with particles of the size of molecular clouds can be, and has been, tried. But the mass and sizes of molecular clouds are not invariant and uniquely defined since they collide and disrupt in typically much less than a galactic rotation. Not only the gas equation of state must be known, but the equation of state of the cloud population must be given (Prendergast 1962, Cowie 1980).

The various approaches to model galactic gas at large scales include Eulerian schemes by finite elements (e.g. Mulder & Liem 1986), "beam" schemes (van Albada et al. 1982), Lagrangian sticky particle models (e.g. Schwarz 1981; Combes & Gerin 1985), or the Smooth Particle Hydrodynamics technique (e.g. Friedli & Benz 1993). None of these techniques with todays computers has the resolving power to describe well a gaseous fractal such as the cold ISM. However, particle techniques have proven efficient because resolution can easily be adapted to local density enhancements. Only the number of particles is presently too small.

Since recent and old observations require a hierarchy with much more levels than previously estimated, clearly hydrodynamics becomes increasingly useless. On the contrary the scale-free and hierarchical aspect missing in the hydrodynamical description should be modelled in some way. Recently Houlahan & Scalo (1992) have built static hierarchical models with up to three levels. A variation of this are the models of Bregman & Ashe (1991) that mix continuous gas and dense "standard" clouds. Independently of us, Hetem & Lépine (1993) have developed a number of recursive methods for constructing inhomogeneous and fractal cloud models.

## 2.6. Fragmentation and isothermal conditions

Since Hoyle's (1953) paper on fragmentation, it is thought that a Jeans unstable and isothermal (because its cooling time is much shorter than its free-fall time) cloud fragments into a small number $N$ of subclouds, that in turn repeat the process recursively at smaller scale as long as isothermal conditions prevail.

Hoyle recommended $N \approx 5$, but the important point is just the order of magnitude, $N$ should be small, as suggested by numerical simulations (see e.g. Scalo 1985). Based on IRAS data, Houlahan & Scalo (1992) derive a value $N = 8 - 10$ for the Taurus region; they compare the Taurus tree architecture to simulated hierarchies of clouds. In the same region, Cernicharo (1991) finds $N = 3 - 6$, based on CO and extinction data, and with a comparable method.

In fact the strict isothermal conditions of a finite self-gravitating gas cloud with no rotation leads rapidly to the conclusion that in many cases no static equilibrium is possible. For example a strict isolated isothermal gas sphere needs an infinite mass, and having a finite mass it evolves by evaporating its outer parts while shrinking its inner parts. This leads to the gravo-thermal catastrophe (Lynden-Bell & Wood 1968; Katz 1978). Once the density ratio between the edge and the centre exceeds 32.1 an isothermal sphere is unstable. The only stable equilibriums of isothermal gas spheres exist when the outer pressure is large enough, in fact comparable to or larger than the gravitational energy density.

The theoretical question that is still obscure today is to characterise the state of an isothermal and self-gravitating ideal gas (e.g., enclosed in a box) which is Jeans unstable, i.e. its internal thermal energy is much smaller than its gravitational energy. If the gas is strictly isothermal its thermal energy cannot be increased or decreased, dissipation vanishes, yet it cannot remain static due to the Jeans instability. Necessarily a dynamical state follows at a scale larger than the molecule mean free-path. As for gases at the laboratory scale a thermodynamical equilibrium should be reached. However since gravity is a long range force its effect is different from short range molecular forces leading to uniform densities. The scale-free character of gravity should also lead to a scale-free behaviour of the unstable isothermal gas. However this scale-free behaviour cannot extend indefinitely at small-scale. The scale at which fragmentation or scale-free behaviour stops may be determined by the transition from isothermal to adiabatic conditions. The transition isothermal-adiabatic is a critical regime at which temperature fluctuations can be large.

So the fractal nature of the cold ISM appears a posteriori as quite natural. In general, gravity dominates all the other forces at large scale, but large scale structures cool fast. Necessarily if nothing like star formation perturbs isothermal conditions a highly chaotic, dynamical but weakly dissipative regime should follow.

## 2.7. Elementary cloudlets, "clumpuscules"

In a real gas cloud isothermal conditions break down at small scales when the initial collapse time-scale $\tau_{\mathrm{ff}} = 1/\sqrt{G\rho}$ becomes shorter than Kelvin-Helmoltz time-scale $\tau_{\mathrm{KH}}$, which, in almost isothermal conditions, is of the order of the heat energy divided by the Stefan-Boltzmann radiation law

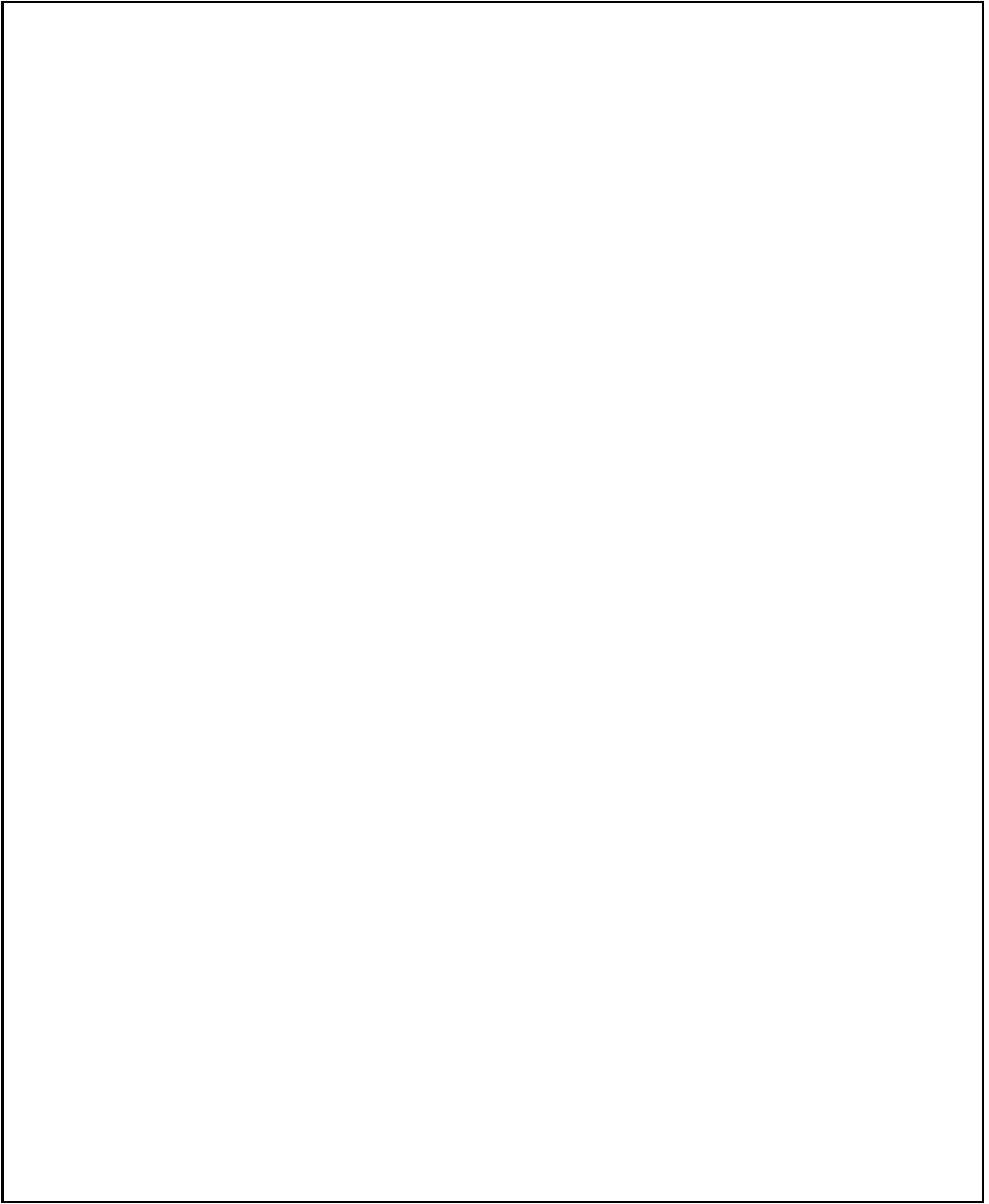

**Fig. 1.** Clumpuscule physical properties as a function of temperature $T$ between 1 and $10^6$ K: mass $M$, radius $R$, virtual luminosity $L$, average density $\rho$, average surface density $\Sigma$, and free-fall time $\tau_{\mathrm{ff}}$ for $\mu = 2.3$ at $T < 3000$ K and $\mu = 0.63$ at $T > 5000$ K for the factor $f = 1$ (solid) and $f = 0.1$ (dashed)

$$\tau_{\mathrm{KH}} = \frac{3}{2} \frac{MkT}{\mu m_{\mathrm{p}}} \Big/ L , \quad \text{where} \quad L = 4\pi f R^2 \sigma T^4 , \tag{2}$$

(Rees 1976; Kippenhahn & Weigert 1991, Chap. 26). Here $\mu$ is the mean molecular weight, $m_{\mathrm{p}}$ the proton mass, $f$ a factor of the order of one, or smaller, depending on departure to spherical geometry and to black-body radiation, and $\sigma$ the Stefan-Boltzmann constant. The equality $\tau_{\mathrm{KH}} = \tau_{\mathrm{ff}}$ yields, with the virial relation $M = 3kTR/G\mu m_{\mathrm{p}}$, a natural minimum mass for fragmentation, that we call below "clumpuscules". By the virial theorem, other main characteristics of these elementary cloudlets can be derived:

$$
\begin{aligned}
M_\bullet &\approx 4 \cdot 10^{-3} & T^{1/4} \, \mu^{-9/4} \, f^{-1/2} & \quad [\mathrm{M}_\odot] &, \\
R_\bullet &\approx 1.5 \cdot 10^2 & T^{-3/4} \, \mu^{-5/4} \, f^{-1/2} & \quad [\mathrm{AU}] &, \\
L_\bullet &\approx 9.2 \cdot 10^{-7} & T^{5/2} \, \mu^{-5/2} & \quad [\mathrm{L}_\odot] &, \\
\rho_\bullet &\approx 1.1 \cdot 10^8 & T^{5/2} \, \mu^{3/2} \, f & \quad [\mathrm{H \, cm}^{-3}] &, \\
P_\bullet &\approx 1.1 \cdot 10^8 & T^{7/2} \, \mu^{1/2} \, f & \quad [\mathrm{K \, cm}^{-3}] &, \\
\Sigma_\bullet &\approx 3.2 \cdot 10^{23} & T^{7/4} \, \mu^{1/4} \, f^{1/2} & \quad [\mathrm{H \, cm}^{-2}] &, \\
\tau_{\mathrm{ff},\bullet} &\approx 9.2 \cdot 10^3 & T^{-5/4} \, \mu^{-3/4} \, f^{-1/2} & \quad [\mathrm{yr}] &,
\end{aligned}
\tag{3}
$$

where $M_\bullet$ is the mass, $R_\bullet$ the radius, $L_\bullet$ the virtual luminosity, $\rho_\bullet$ the average density, $P_\bullet$ the average pressure, $\Sigma_\bullet$ the average surface density, and $\tau_{\mathrm{ff}}$ the free-fall (or dynamical) time. By construction $\tau_{\mathrm{KH}} = \tau_{\mathrm{ff}}$.

At $T = 3$ K and for $0.1 \leq f \leq 1$ the clumpuscules typical parameters are:

$$
\begin{aligned}
M_\bullet &\sim 0.8 - 2.7 \cdot 10^{-3} & \quad [\mathrm{M}_\odot] &, \\
R_\bullet &\sim 23 - 73 & \quad [\mathrm{AU}] &, \\
L_\bullet &\sim 1.8 \cdot 10^{-6} & \quad [\mathrm{L}_\odot] &, \\
\rho_\bullet &\sim 0.6 - 6 \cdot 10^9 & \quad [\mathrm{H \, cm}^{-3}] &, \\
P_\bullet &\sim 0.5 - 5 \cdot 10^9 & \quad [\mathrm{K \, cm}^{-3}] &, \\
\Sigma_\bullet &\sim 0.8 - 2.7 \cdot 10^{24} & \quad [\mathrm{H \, cm}^{-2}] &, \\
\tau_{\mathrm{ff},\bullet} &\sim 1.2 - 3.9 \cdot 10^3 & \quad [\mathrm{yr}] &.
\end{aligned}
\tag{4}
$$

The run of some of these quantities are given in Fig. 1 for $\mu = 2.3$ at $T < 3 \cdot 10^3$ K (neutral $H_2$ and He), and $\mu = 0.63$ at $T > 5 \cdot 10^3$ K (ionised H and neutral He), for $f = 0.1$, and $f = 1$. The following comments can be made:

1. The clumpuscule typical mass is weakly dependent on temperature. The fact that $M_\bullet$ is of the order of a Jupiter at low $T$ or a brown dwarf at higher $T$ does not imply that such clumpuscules resemble stars or Jupiters. As $T$ increases the mass *increases*, and the size *decreases*, leading eventually to denser and hotter stellar-like objects. The decreasing of the virial radius $R_\bullet$ does not mean that all the mass contract to smaller radii; on the contrary, we expect that, as the mass contract in average, *a part* of the mass expands and evaporates to larger radii, similarly to red giants or globular clusters. The decreasing of $\mu$ at $T > 2 \cdot 10^3$ K due to $H_2$ dissociation and H ionisation increases $M_\bullet$ by a factor $(2.3/0.63)^{9/4} \approx 18$.

2. Although $L$ is explicitly defined as depending on $f$ in Eq. (2), this dependency cancels through the $R$ and $T$ dependencies. The luminosity given here is virtual since the medium is thought to be nearly isothermal; within a factor of order unity $L_\bullet$ corresponds to the gravitational power $L_{\mathrm{gra}} \approx v^5/G$ mentioned in Paper I. It is also discussed below (Sect. 3.3).

3. If made of primordial gas, at low $T$ the clumpuscules are almost completely transparent in the optical wavelengths since the average column density corresponds to only a few grams of H, $H_2$, and He per cm$^2$. However we expect that the clumpuscules are themselves inhomogeneous, approximately isothermal. Thus the average density $\rho_\bullet$ and surface density $\Sigma_\bullet$ are only rough indicators of the actual clumpuscule opacity. However clumpuscules might be opaque also in the optical if some fraction of $H_2$ freezes into $H_2$ snow (see Setc. 6).

4. Too small masses have an increasingly *long* cooling time. We will see in the next Section that with an overall fractal organisation clumpuscules collide supersonically with others as fast as they contract in the adiabatic phase. This prevent collapses. Too large masses are Jeans unstable, fragment and cool faster than they can collapse. A clumpuscule is never in a really static state.

## 3. Scaling relations in fractal clouds

### 3.1. Clump definition

Here a clump is defined as a finite amount of mass $M$ distributed not necessarily in a finite volume, but characterised by a finite scale-length $r_L$. A clump is made of sub-clumps distributed according to a probability density law $\rho(r, r_L)$.

Since fragmentation following a Jeans instability in a gravitating medium is a chaotic process, there is no reason to expect that a fragmenting cloud behaves exactly self-similarly. Instead we expect that it behaves in a statistically self-similar way. In average it fragments into subclumps following a probability distribution of their number, mass and size.

So a hierarchical structure can be defined in a recursive way. Simplifying over the real situation here, a hierarchical clump of level $L > 0$ with the scale-length $r_L$ is made of $N$ independent subclumps themselves hierarchical clumps of level $L - 1$ with the scale-length $r_{L-1}$. The subclumps are distributed randomly according to a density distribution $\rho(r, r_L)$. When $L = 0$ the recursion stops, a $L = 0$ clump represents the most basic mass unit ("atom") of the whole hierarchy.

In this paper only the distribution of subclumps follows a probability distribution. Of course we could also determine $N$ and $r_L$ by probability distributions. Also we could define clumps according to non-spherical distributions. We let these complications for further studies.

Since clumps may have mass distributed outside their own scale-length $r_L$, this outer mass can be considered as the left-over mass following a fragmentation process. Reasonable mass distributions should have a sufficiently large fraction of the mass within the scale-length $r_L$ (as e.g. the Plummer distribution $\rho \sim (r^2 + r_L^2)^{-5/2}$, and a negligible mass at the scale of the parent cloud, otherwise the hierarchical model is not very useful. Scalo (1985) defines a fragmentation efficiency parameter, separating the fragment densest part at $r < r_L$ from the matter outside $r_L$. Obviously if the fragmentation efficiency is not very high, say 80% per level, after only a few levels the mass included in the smallest fragments becomes negligible, and the hierarchical model less relevant. Here we prefer to include this free parameter directly into the clump density distribution $\rho$, which is an unknown function yet.

### 3.2. Scaling relations

#### 3.2.1. Fractal dimension of a mass distribution

Let $M$ be the mass within a radius $r$, $M(r) = 4\pi \int_0^r \rho(s)s^2 \, ds$. If the mass obeys the self-similar scaling relation

$$\frac{M}{M_0} = \left(\frac{r}{r_0}\right)^D ,\qquad (5)$$

for a reference mass $M_0$ at the scale $r_0$, then $D$ defines the fractal dimension of the mass distribution (Mandelbrot 1982).

Otherwise stated

$$D = \frac{\log(M/M_0)}{\log(r/r_0)} ,\qquad (6)$$

i.e. the knowledge of the largest and the smallest scales and masses of a cloud allows to determine a global fractal dimension $D$. For example if the total mass of a typical molecular cloud is $5 \cdot 10^5 \, \mathrm{M_\odot}$ at a scale of 30 pc, as inferred from the average surface density of $170 \, \mathrm{M_\odot} \, \mathrm{pc}^{-2}$ determined by Solomon et al. (1987), and the smallest clouds would be clumpuscules with a mass of $0.001 \, \mathrm{M_\odot}$ for a scale of 30 AU, the global fractal dimension would be $D = 1.64$.[2]

#### 3.2.2. Mass-number relation

Since the mass $M_L$ of a clump at level $L$ consists of $N$ sub-clumps of mass $M_{L-1}$ at level $L-1$, it follows

$$
\begin{aligned}
M_L = N M_{L-1} &\Rightarrow r_L^D = N r_{L-1}^D \\
&\Rightarrow \alpha \equiv \frac{r_{L-1}}{r_L} = N^{-1/D} .
\end{aligned}\qquad (7)
$$

We deduce that if the scale ratio $\alpha$ is constrained by a maximum value $\alpha_{max}$, then $N > \alpha_{max}^{-D}$. In order to meet a minimum contrast between levels, $N$ has to be much larger at $D = 3$ than at $D = 1$.

For the moment, we keep generality in allowing $N$ to depend on $L$, since several interesting properties depend weakly on a constant $N$ that a strict self-similar distribution would have. Indeed the $M_L - N$ relation is independent of $D$ (see e.g. Scalo 1985, p. 243)[3].

---

[2] Applying Eq. (6) to the Galaxy and its stars as the largest and smallest clumps yields $D \approx 1$. Similar application to large scale structures in the Universe yields $D \approx 1.2$ (cf. de Vaucouleurs 1970; Mandelbrot 1982; Coleman & Pietronero 1992).

[3] The empirically determined relation $N(M) \sim 1/M^{1+x}$, $x \approx 1.1 - 1.7$ (see e.g. Stutzki & Güsten 1990; Genzel 1992) is a distribution of clump masses. It implicitly assumes that at least a fraction of the mass of each clump does not belong to the subclumps. The definition of a clump needs therefore a *smooth* mass model, such as a Gaussian law. Consequently the resulting mass spectrum is directly dependent on the clump mass model.

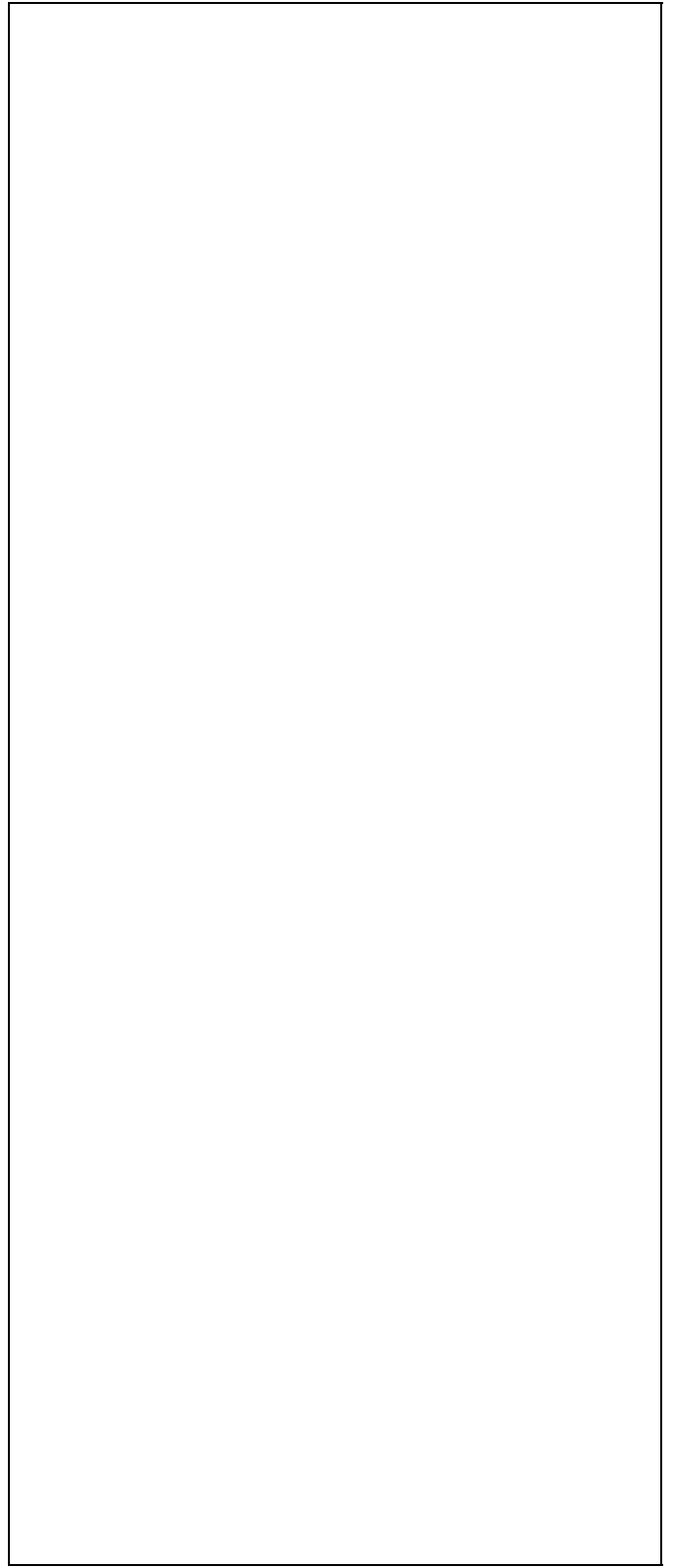

**Fig. 2.** A fractal model with 4 levels and 5 clumps per levels with from top to bottom $D = 1.5$, 2, and 3. The lowest level clumps are shown as opaque balls. The circles show the extensions of the higher level clumps. The background grid suggests a finite resolution device

So keeping generality we note $N_L$ instead of $N$, and we define

$$n_L \equiv \prod_{i=1}^{L} N_i \ . \tag{8}$$

The symbol $n_L$ is the total number of elementary clouds contained at the level $L$. If $N_L = N$ is constant, $n_L = N^L$. It follows from this definition that

$$M_L = n_L M_0 \ , \qquad \frac{r_L}{r_0} = n_L^{1/D} \ . \tag{9}$$

### 3.2.3. Average density and surface density

The average density $\langle \rho_L \rangle$ of a cloud at level $L$ scales as

$$\frac{\langle \rho_L \rangle}{\langle \rho_0 \rangle} = \frac{M_L/M_0}{(r_L/r_0)^3} = n_L^{1-(3/D)} = \left( \frac{r_L}{r_0} \right)^{D-3} \ , \tag{10}$$

so when $D < 3$ density always increases at small scale.

Similarly, the average surface density $\langle \Sigma_L \rangle$ of a *single* cloud at level $L$ scales as

$$\frac{\langle \Sigma_L \rangle}{\langle \Sigma_0 \rangle} = \frac{M_L/M_0}{(r_L/r_0)^2} = n_L^{1-(2/D)} = \left( \frac{r_L}{r_0} \right)^{D-2} \ . \tag{11}$$

If $D > 2$, $\langle \Sigma_L \rangle$ is larger at large scale than at small scale, while if $D < 2$, it is the opposite. When $D < 2$ the probability that several clouds at the same level overlap in projection decreases. For a *uniform* probability of coverage by subclouds, the ratio of the surfaces $(r_{L-1}/r_L)^2 = \alpha^2$ and the probability $P(n)$ to have $n$ subclouds at the same position among a collection of $N$ subclouds is given by a binomial distribution, since the probability to have one single subcloud at a given position is $\alpha^2$:

$$P(n) = \binom{N}{n} \left( \alpha^2 \right)^n \left( 1 - \alpha^2 \right)^{N-n}$$
$$= \binom{N}{n} \left( N^{-2/D} \right)^n \left( 1 - N^{-2/D} \right)^{N-n} \ . \tag{12}$$

It can be verified that the probability to have more than one superposed subclouds is larger at larger $D$ and larger $N$. By decreasing $D$ a better view through the fractal is possible, but the subclouds are then denser. In addition to the fact that in realistic clouds the coverage probability is varying over the cloud, the above probability cannot be extended simply down to the lowest level of the hierarchy because the subclouds positions along the hierarchy are *correlated*. This complicates greatly the probabilistic evaluation, and encourages investigating the question with a Monte-Carlo approach. Fig. 2 illustrates what happens when $D$ varies from 1.5 to 3 in a hierarchical cloud model where the probability of presence of a single subclump decreased with $r$ (model "A" described below). When $D$ decreases more clumps are directly visible, but rarely completely isolated. Since each ball has the same mass, their surface density is much higher at low $D$.

### 3.2.4. Filling factor

Neglecting the eventual overlapping of clumps in space (a good approximation at $D \ll 3$), the volume filling factor $f_L$ of the elementary clumps at level 0 at the scale of level $L$ is simply the ratio of the volumes,

$$f_L = \frac{n_L r_0^3}{r_L^3} = n_L^{1-(3/D)} = \left( \frac{r_0}{r_L} \right)^{3-D} \ , \tag{13}$$

so is rapidly very small if $D < 3$ and $L$ is large. In hierarchical structures the filling factor is a meaningless concept unless the largest and smallest levels are indicated.

### 3.3. Properties of virialised fractal clouds

#### 3.3.1. Virial velocity

In a virialised and sufficiently hierarchised mass distribution (say $\alpha < 0.2$), the gravitational energy of a clump at level $L$ can be approximated by its own gravitational energy, neglecting the energy contributions of the clumps at other levels. Incidentally this is precisely the reason allowing a clump to keep its identity. This approximation (which implies that the mass fraction outside $r_L$ is small) allows to use the virial theorem at each level of the hierarchy, giving a relation with the virial velocities

$$v_L^2 \equiv \frac{G M_L}{r_L} = v_0^2 \left( \frac{r_L}{r_0} \right)^{D-1} \ . \tag{14}$$

So knowing the typical velocities and sizes of the largest and smallest clumps allows to estimate $D$:

$$D = 1 + 2 \, \frac{\log(v_L/v_0)}{\log(r_L/r_0)} \ . \tag{15}$$

The exponent $\kappa$ of the Larson relation $v \sim r^\kappa$ in the ISM (for a review see e.g. Elmegreen 1992) ranges between 0.3 and 0.5, which suggests by Eq. (15) that in the ISM, $1.6 < D < 2$.

The approximation of virial equilibrium made here is probably the most critical one, because it neglects the outer clump gravitation, and other physical effects such as the interclump gas or magnetic pressure. Furthermore the virial assumption supposes that the second time-derivative of the inertia tensor $I$ vanishes, which strictly should be exceptional in a turbulent hierarchical gas. Also a more complete virial assumption would include internal and external pressures $P_{\text{int}}$ and $P_{\text{ext}}$,

$$\tfrac{1}{2} \ddot{I} = 2 E_{\text{kin}} + E_{\text{gra}} + 3 P_{\text{int}} V - 3 P_{\text{ext}} V \ , \tag{16}$$

where $I$ is the moment of inertia. Since the volume $V$ is the same in both pressure terms of the equation, if these pressures are similar their contribution remains small.

#### 3.3.2. Dynamical time

Keeping the simplest virial assumption Eq. (14), and using also Eqs. (5), (7), the mean dynamical (or crossing) time $\tau_{\text{dyn}, L}$ at level $L$ is given by

$$\tau_{\text{dyn}, L} \equiv \frac{r_L}{v_L} = \tau_{\text{dyn}, 0} \left( \frac{r_L}{r_0} \right)^{(3-D)/2} \ . \tag{17}$$

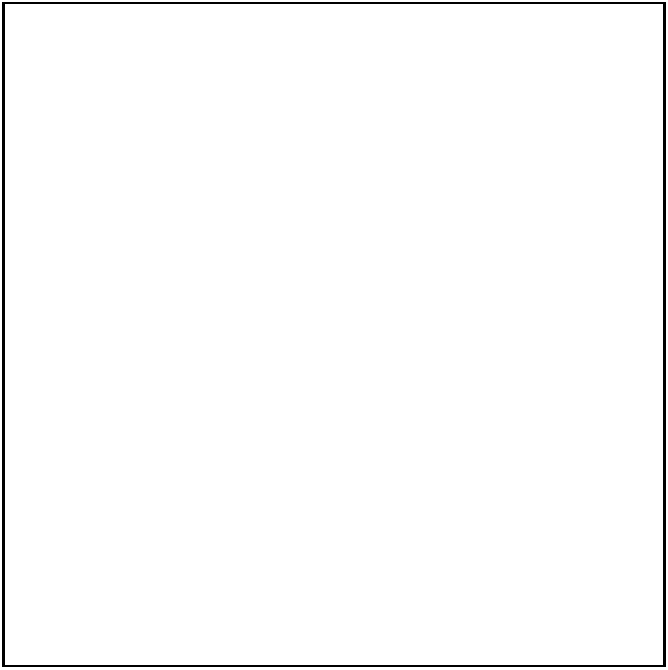

**Fig. 3.** Ratios $C_{\mathrm{int}}$ (dash) and $C_{\mathrm{ext}}$ (solid) in the $N - D$ diagram for $2 < N < 10^3$

The dynamical time is independent of the level for $D = 3$, and it decreases at small scales when $D < 3$.

### 3.3.3. Collision time

Similarly, the mean collision time $\tau_{\mathrm{col}, L}$ at level $L$, using Eqs. (7), (17), is given by

$$\tau_{\mathrm{col}, L} \equiv \left[ \frac{N_L}{\frac{4}{3} \pi r_L^3} \, \pi (2 r_{L-1})^2 v_L \right]^{-1} = \frac{\tau_{\mathrm{dyn}, L}}{3 N_L} \left( \frac{r_L}{r_{L-1}} \right)^2 . \quad (18)$$

So the ratio of $\tau_{\mathrm{col}, L}$ to $\tau_{\mathrm{dyn}, L}$ is

$$C_{\mathrm{int}} \equiv \frac{\tau_{\mathrm{col}, L}}{\tau_{\mathrm{dyn}, L}} = \frac{1}{3} N_L^{(2/D)-1} . \quad (19)$$

For $D > 2$ the ratio $C_{\mathrm{int}}$ is always less than 1, so $D > 2$ is unfavourable for maintaining a fragmented cloud because it tends to dissolve due to *internal* collisions. On the contrary for $D < 2$ and large $N_L$ this ratio exceeds 1, and the collisions are rare enough to allow the $N_L$ subclumps to form a clump for a time much longer than its dynamical time.

The ratio of the collision time to the dynamical time at the adjacent lower level indicates if subclumps collide frequently with respect to their own dynamical time. If external collisions occur over a time-interval shorter than the dynamical time, a new virial equilibrium can hardly be reached between collisions and the clump should disrupt. This ratio is given by

$$C_{\mathrm{ext}} \equiv \frac{\tau_{\mathrm{col}, L}}{\tau_{\mathrm{dyn}, L-1}} = \frac{1}{3} N_L^{(7-3D)/(2D)} . \quad (20)$$

For $D > 7/3$ this ratio is always smaller than 1, so $D > 2.33$ is unfavourable for maintaining a fragmented cloud because *external* collisions tend to dissolve it.

Fig. 3 shows the ratios $C_{\mathrm{int}}$ and $C_{\mathrm{ext}}$ in a $N - D$ diagram. The system is much less collisional at low $D < 2$ and large $N$. At small $N \lesssim 10$ and $D > 1$ the system is always collisional, but much less collisional at $D \approx 1 - 2$ than at $D > 2$. This is crucial for understanding how clumpuscules can survive without collapsing.

### 3.3.4. Mach number and shock conditions

The collision strength is characterised by the Mach number $\mathcal{M}_L$,

$$\mathcal{M}_L \equiv \frac{v_L}{v_{L-1}} = \left( \frac{r_L}{r_{L-1}} \right)^{(D-1)/2} = N_L^{[1-(1/D)]/2} . \quad (21)$$

The collisions are "supersonic" at $D > 1$, and "subsonic" at $D < 1$, because the collision speed is respectively larger or smaller than the clump internal subclump velocities. In the range $1 < D < 2$ the collisions are only slightly supersonic, $\mathcal{M}_L < 2$ for $N < 16$.

Let us examine a consequence of having clump "supersonic" motion. At any level we have $N_L$ clumps the mass of which is concentrated mainly within the scale-length $r_L$. The residual mass exterior to $r_L$ interacts "supersonically" with other clumps. We recall that the interclump medium is also thought to be made of subclumps, not of pure smooth gas. In front of each clump, at a distance of the order of $r_L$ a shock in the subclump population can be expected. Treating this subclump population as a gas of adiabatic index $\gamma$, we can apply the shock conditions, i.e. the requirement of mass and momentum conservation. This leads to the ratio between the pre- and post-shock densities:

$$\frac{\rho_{\mathrm{post}}}{\rho_{\mathrm{pre}}} = \left[ \frac{2}{\gamma + 1} \frac{1}{\mathcal{M}^2} + \frac{\gamma - 1}{\gamma + 1} \right]^{-1} , \quad (22)$$

(see e.g. Elmegreen 1992). In a self-similar structure $\mathcal{M}$ is given by Eq. (21), so $\mathcal{M}$ depends only on $N_L$ and $D$. The ratio $\rho_{\mathrm{post}} / \rho_{\mathrm{pre}}$ depends furthermore on $\gamma$.

Now the ratio of the average clump densities in adjacent levels is given by Eq. (10), also a simple expression of $N$ and $D$. At the transition region between the inner part of a clump at $r < r_L$ and its outer part at $r > r_L$, the ratio between the inner and outer density is only a fraction less than unity of $\langle \rho_{L-1} \rangle / \langle \rho_L \rangle$. If we require that the density ratio $\rho_{\mathrm{post}} / \rho_{\mathrm{pre}}$ in Eq. (22) matches a fraction $\eta \lesssim 1$ of the density contrast,

$$\eta \frac{\langle \rho_{L-1} \rangle}{\langle \rho_L \rangle} = \frac{\rho_{\mathrm{post}}}{\rho_{\mathrm{pre}}} , \quad (23)$$

we obtain after some algebra a quartic equation for $x \equiv N^{1/D}$

$$2 x^4 + N(\gamma - 1) x^3 = \frac{N^2}{\eta} (\gamma + 1) , \quad (24)$$

which can be solved by standard means. Only one of the four solutions is real for $D$ in the relevant domain $N \geq 2$, $\gamma \geq 1$, and $0 < \eta \leq 1$. The solution for $D$ is slowly varying in $N$

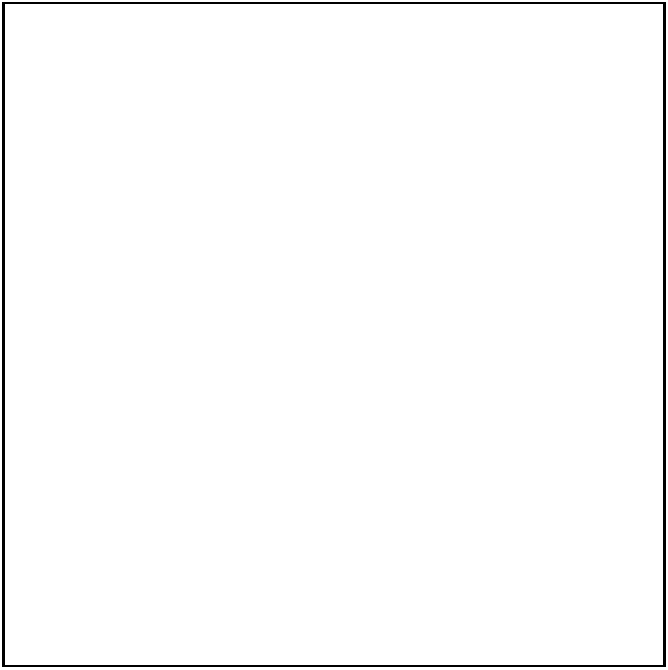

**Fig. 4.** Real solutions for $D$ of the quartic equation (24) for $\eta = 0.3$ (bottom), 0.5 (middle), and 1 (top) in a $(N - \gamma - D)$ diagram

and $\gamma$, and in the range $D \leq 3$. For $\gamma = 1$ we have the simple solution

$$D = \frac{2}{1 - \frac{1}{2}\frac{\ln \eta}{\ln N}} , \qquad (25)$$

which shows that in this case $D \leq 2$. For $\eta = 0.5$, $D$ varies from $D = 1.52$ at $N = 3$ to $D = 1.79$ at $N = 20$, to $D = 2$ at $N = \infty$.

We have also:

$$\lim_{\gamma \to \infty} D = \frac{3}{1 - \frac{\ln \eta}{\ln N}} , \quad \text{and} \quad \lim_{N \to \infty} D = 3 , \qquad (26)$$

So a consistent self-similar regime between the fragmentation process and the shock conditions can be established over a wide range of scales $\gamma$, $N$ and $\eta$. It is found that for plausible parameters (e.g. around $\gamma = 5/3$, $N = 10$ and $\eta = 0.5$) then the fractal dimension $D$ is constrained to take values around $1.6 - 2$, so similar to the $D$ derived from Larson's relations.

### 3.3.5. Reynolds number

The "viscosity" between clumps at adjacent levels is characterised by the Reynolds number $\mathcal{R}_L$, i.e. the ratio of the viscosity coefficients $\nu_l = v_L r_L$:

$$\mathcal{R}_L \equiv \frac{\nu_L}{\nu_{L-1}} = \left(\frac{r_L}{r_{L-1}}\right)^{(D+1)/2} = N_L^{[1+(1/D)]/2} . \qquad (27)$$

So either decreasing $D$ or increasing $N_L$ increases the Reynolds number, i.e. decreases viscosity. In a fractal gas spanning a large range of scales the overall viscosity (transfer of momentum at different scales) can be very low.

### 3.3.6. Angular momentum

Assuming a constant fraction of rotational energy along the levels, $v_{\mathrm{rot}, L}/v_L = \text{constant}$, the specific angular momentum $J_L$ scales as $r_L v_L$, so the ratio $J_L/J_0$ scales as the Reynolds number,

$$\left(\frac{J_L}{J_0}\right) = \left(\frac{r_L}{r_0}\right)^{(D+1)/2} . \qquad (28)$$

Thus for positive $D$ most of the angular momentum has to be in large scale clumps. Angular momentum can hardly diffuse down the hierarchy because velocity would then increase as $r^{-1}$ and the maximum rotation velocity $v_{\mathrm{rot}, L} = v_L$ would soon be exceeded. The clumps would then be unstable and would eject the excess angular momentum. The extreme situation is reached when at all levels the clumps have received the maximum amount of angular momentum compatible with a stable configuration.

### 3.3.7. Kinetic energy density

The kinetic energy density $u_L = \frac{1}{2}\rho_L v_L^2$ corresponds to a kinetic pressure and scales as

$$\frac{u_L}{u_0} = \frac{\rho_L v_L^2}{\rho_0 v_0^2} = \left(\frac{r_L}{r_0}\right)^{2(D-2)} . \qquad (29)$$

So $u_L$ is large at large scale when $D > 2$, and large at small scale when $D < 2$. When $D = 2$, $u$ is constant. A fragmented structure is harder to dislocate if the kinetic pressure increases at small scale, therefore $D < 2$ is favourable for maintaining a fractal.

### 3.3.8. Total velocity dispersion

The total kinetic energy $E_{\mathrm{kin}, L}$ is the sum of the kinetic energies along the hierarchy, and is clearly an increasing and diverging function of the number of levels. Up to the level $L$ the total velocity dispersion squared $\sigma_L^2$ ($\sigma_0 \equiv v_0$), proportional to the kinetic energy per unit mass, is (using Eqs. (9), (14)),

$$\frac{\sigma_L^2}{\sigma_0^2} = \frac{E_{\mathrm{kin}, L}}{\frac{1}{2}n_L M_0 v_0^2} = \sum_{l=0}^{L} \underbrace{\frac{n_{L-l} M_L}{n_L M_0}}_{=1} \frac{v_l^2}{v_0^2}$$
$$= \sum_{l=0}^{L} \left(\frac{r_l}{r_0}\right)^{D-1} = \sum_{l=0}^{L} n_l^{1-(1/D)} . \qquad (30)$$

If $N_L = N$ is a constant, then $n_L = N^L$, and this ratio reduces to

$$\frac{\sigma_L^2}{\sigma_0^2} = \frac{\left(N^{1-(1/D)}\right)^{L+1} - 1}{N^{1-(1/D)} - 1} . \qquad (31)$$

Since $L = D \ln(r_L/r_0)/\ln N$ the velocity dispersion ratio becomes

$$\frac{\sigma_L^2}{\sigma_0^2} = \frac{(r_L/r_0)^{D-1} N^{1-(1/D)} - 1}{N^{1-(1/D)} - 1} . \qquad (32)$$

**Table 1.** Relation between the velocity dispersion ratio $\sigma_L/\sigma_0$, the size ratio $r_L/r_0$, $N$ and $D$ in a virialised fractal structure (Eq. (32))

| $N=3$ | | | $N=5$ | | | $N=10$ | | | $N=\infty$ | | |
|---|---|---|---|---|---|---|---|---|---|---|---|
| $\sigma_L/\sigma_0$ | $r_L/r_0$ | $D$ | $\sigma_L/\sigma_0$ | $r_L/r_0$ | $D$ | $\sigma_L/\sigma_0$ | $r_L/r_0$ | $D$ | $\sigma_L/\sigma_0$ | $r_L/r_0$ | $D$ |
| $10^1$ | $10^4$ | 1.35 | $10^1$ | $10^4$ | 1.39 | $10^1$ | $10^4$ | 1.43 | $10^1$ | $10^4$ | 1.50 |
| $10^1$ | $10^6$ | 1.21 | $10^1$ | $10^6$ | 1.24 | $10^1$ | $10^6$ | 1.26 | $10^1$ | $10^6$ | 1.33 |
| $10^2$ | $10^4$ | 1.90 | $10^2$ | $10^4$ | 1.93 | $10^2$ | $10^4$ | 1.96 | $10^2$ | $10^4$ | 2.00 |
| $10^2$ | $10^6$ | 1.59 | $10^2$ | $10^6$ | 1.61 | $10^2$ | $10^6$ | 1.63 | $10^2$ | $10^6$ | 1.67 |
| $10^3$ | $10^4$ | 2.42 | $10^3$ | $10^4$ | 2.45 | $10^3$ | $10^4$ | 2.47 | $10^3$ | $10^4$ | 2.50 |
| $10^3$ | $10^6$ | 1.94 | $10^3$ | $10^6$ | 1.96 | $10^3$ | $10^6$ | 1.97 | $10^3$ | $10^6$ | 2.00 |

We see that for $N^{1-(1/D)} \gg 1$ and $D > 1$, $(\sigma_L/\sigma_0) \approx (r_L/r_0)^{(D-1)/2}$, so is weakly dependent on $N^{1-(1/D)}$, and tends toward a Larson type velocity-size relation. Straightforward calculations show that for relevant values of $\sigma_L/\sigma_0$ and $r_L/r_0$ for the cold ISM, i.e. $10 \lesssim \sigma_L/\sigma_0 \lesssim 10^3$, and $10^4 < r_L/r_0 < 10^6$, then $1.21 \lesssim D \lesssim 2.50$, for any $N \geq 3$ (cf. Table 1.).

### 3.3.9. Power and energy dissipation

The typical power $P_L$ that *might* be exchanged at level $L$ is of the order of the gravitational energy (twice the kinetic energy) divided by the dynamical time. Therefore, using Eq. (14),

$$P_L = M_L v_L^2 \frac{v_L}{r_L} = \frac{v_L^5}{G} . \tag{33}$$

Since virial equilibrium is assumed, $P_L$ is a virtual power that could be delivered if equilibrium would be perturbed. This virtual power scales as

$$\frac{P_L}{P_0} = \left(\frac{v_L}{v_0}\right)^5 = \left(\frac{r_L}{r_0}\right)^{\frac{5}{2}(D-1)} . \tag{34}$$

So for $D > 1$ the power increases at large scale.

The specific power (power per unit mass) $p_L \sim v_L^3/r_L$ is an interesting quantity used in discussions about the turbulence in the ISM (Fleck 1981, 1983). The specific power scales as

$$\frac{p_L}{p_0} = \frac{P_L/P_0}{M_L/M_0} = \left(\frac{r_L}{r_0}\right)^{(3D-5)/2} . \tag{35}$$

So the specific power increases at small scales when $D < 5/3$, and vice versa when $D > 5/3$. We have the interesting regime of constant specific power at $D = 5/3 \approx 1.67$.

## 4. Fractal cloud models

Here we describe the properties of static fractal clouds generated by Monte-Carlo simulations. In this paper we don't show that our models are dynamically stable in average, or that our models are compatible with the ISM observations. As explained in Sect. 2, our inability to treat correctly the radiation transfer equation in a fractal medium, and our ignorance of the true 3D distribution in the ISM prevent us to set up immediately adequate models. Our first aim here is to illustrate with particular models that important observational differences occur in fractal models with $D < 3$ with respect to conventional smooth cloud models with $D = 3$.

### 4.1. Density laws

The centers of the clouds and subclouds are distributed randomly according to a 3D density law $\rho(r, r_L)$, where $r_L$ is the scale-length. For the sake of simplicity here the density law $\rho$ is kept spherical and identical at all levels, except that all length scales between two levels have a fixed ratio $\alpha \equiv r_{L-1}/r_L$. Clearly complications, i.e. variations depending on the level, can easily be introduced. The cloud projection properties are studied as the fractal dimension and other parameters are varied.

The average clump distribution of the models is built to resemble an isothermal distribution at all levels. In a strict isothermal mass distribution most of the mass lies at large distances ($M \sim r$). If the medium is collisional, the subclumps must be significantly truncated in the outer parts, i.e. the density law must decrease faster than $r^{-2}$ in the outer parts. In order to be roughly in agreement with the above physical constraints, the following average clump density laws have been considered:

Density A: $\quad \rho(r, r_L) \sim \begin{cases} \dfrac{1}{(r/r_L)^2} & \text{for } r < r_L , \\ 0 & \text{for } r \geq r_L , \end{cases} \tag{36}$

an abruptly truncated singular isothermal sphere. This model is not completely realistic as fragmentation is unlikely to be 100% efficient, and collisions and tidal interactions partly dislocate clumps. However this extreme model is useful as comparison with the next one:

Density B: $\quad \rho(r, r_L) \sim \dfrac{1}{(r/r_L)^2} \dfrac{1}{[1+(r/r_L)]^5} , \tag{37}$

a less abruptly truncated isothermal sphere. Here $1/16 \approx 6\%$ of the total mass is exterior to $r_L$, so fragmentation is efficient at 94%. Recall that these smooth density laws can be viewed as the probability density of subclump presence. If $N$

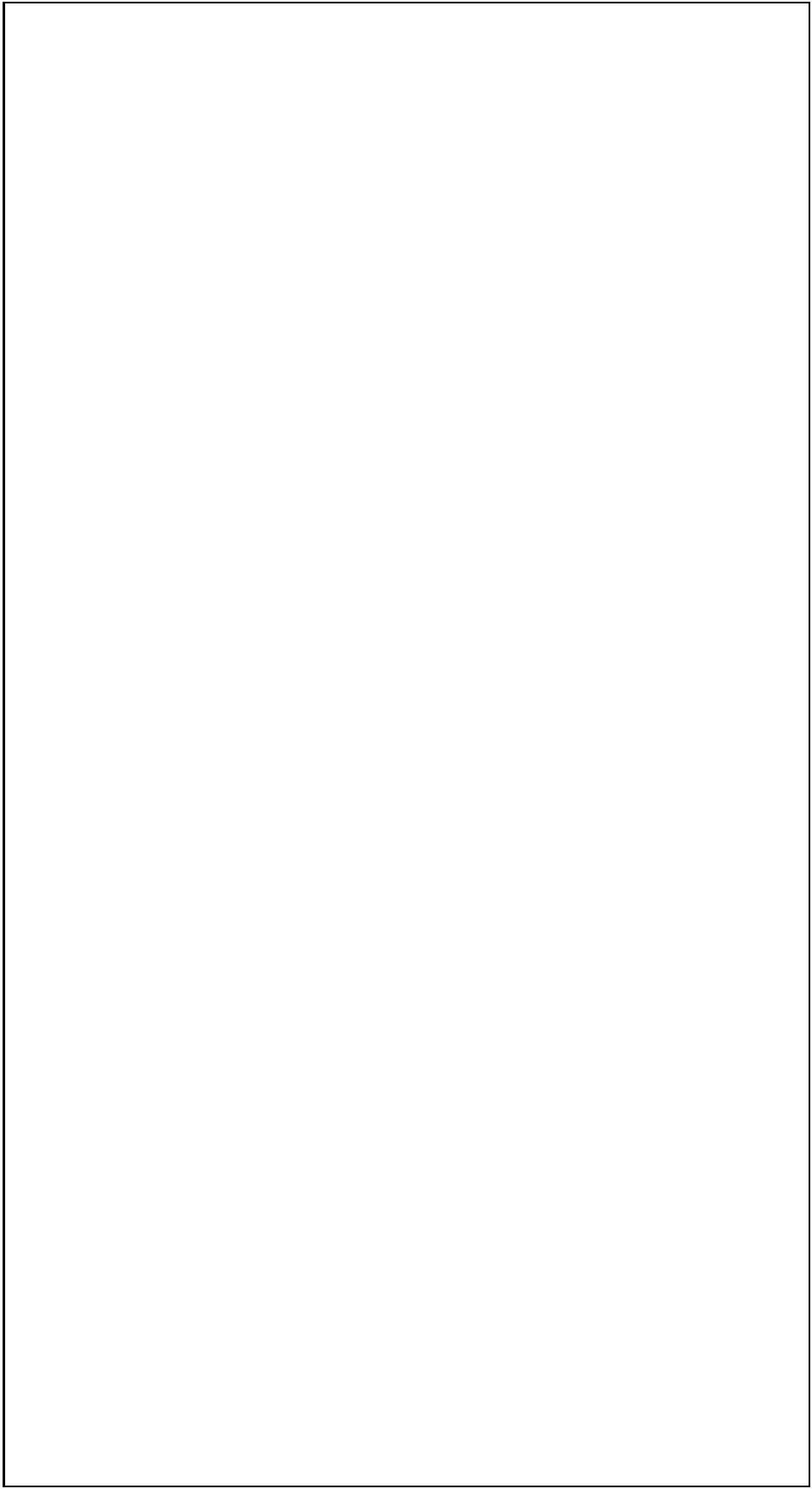

**Fig. 5.** Projected mass distributions (in log) of hierarchical clouds with density laws A (top) and B (bottom). The intensity bar covers 15 magnitudes. The model parameters are $N = 10$, $L = 9$, and $D = 3$, $D = 2.5$, $D = 2$, and $D = 1.5$. The main effect of increasing $N$ is to augment the overall spherical symmetry

is small the effective mass distribution is then far from being smooth and spherically symmetric, and the precise form of $\rho$ is less important.

In order to distribute points randomly in $r$ according to the above density laws from a uniform random distribution $\xi$ between $[0, 1[$, we calculate the inverse function of

$$\xi(r) = \int_0^r 4\pi s^2 \rho(s, r_L)\, \mathrm{d}s \Big/ \int_0^\infty 4\pi s^2 \rho(s, r_L)\, \mathrm{d}s \ , \tag{38}$$

which yields:

$$\begin{aligned} \text{A:} \quad & r(\xi) = r_L \xi \ , \\ \text{B:} \quad & r(\xi) = r_L \left( (1-\xi)^{-1/4} - 1 \right) \ . \end{aligned} \tag{39}$$

A uniform random distribution over a sphere is determined for the polar angles $\phi$ and $\theta$ by drawing $\phi$ uniformly over $[0, 2\pi[$, and uniformly $\cos\theta$ over $[-1, 1]$.

### 4.2. Calculations

Hierarchical models are generated most easily with a language–compiler allowing *recursive programming*. In this work we use the standard language Fortran-77 with the recursion extension of Sun-Sparc computers. Recursive algorithms can often be replaced by more efficient sequential algorithms, but a recursive approach greatly facilitates changes of the model or of parameters.

The length scale $r_L$ at the highest level is set to 1. Then the *centers* of the $N$ subclumps are randomly drawn according to the density laws A or B. The clump centers at level $L$ are added to the parent center at level $L + 1$. Then, for each center, the same algorithm is recursively called with a new length scale $r_{L-1} = \alpha r_L$, and this down to the level $L = 1$. At level $L = 1$ the recursion is stopped, and instead $N$ elementary masses are projected onto a square grid.

So the 3D distribution is not stored, but simultaneously to the cloud calculation the 3D mass distribution is projected along one direction on a square array of fixed resolution, allowing to reach a much higher resolution ($2048^2$) in the projection grid than e.g. Hetem and Lépine (1993) who reach a resolution of $64^3$. The resulting image is made of a superposition of cloudlets with an effective size equal to one pixel. For given $D$ and $N$ the minimum number of levels $L$ is

$$L \gtrsim D \frac{\log \frac{1}{2} 2048}{\log N} \ . \tag{40}$$

The only effect of taking $L$ much larger than the equality is to waste computer time, because then the low level clumps cannot be resolved. A cloud with $10^9$ cloudlets can be generated within a few hours on a contemporary fast workstation.

In order to study resolution effects, the lower resolution images with $1024^2, 512^2, \ldots, 32^2$ pixels are also simply derived by summation of $2 \times 2$ pixel blocks in the twice as large image.

The mass of a single cloudlet averaged over one pixel defines the smallest possible surface density present in the grid.

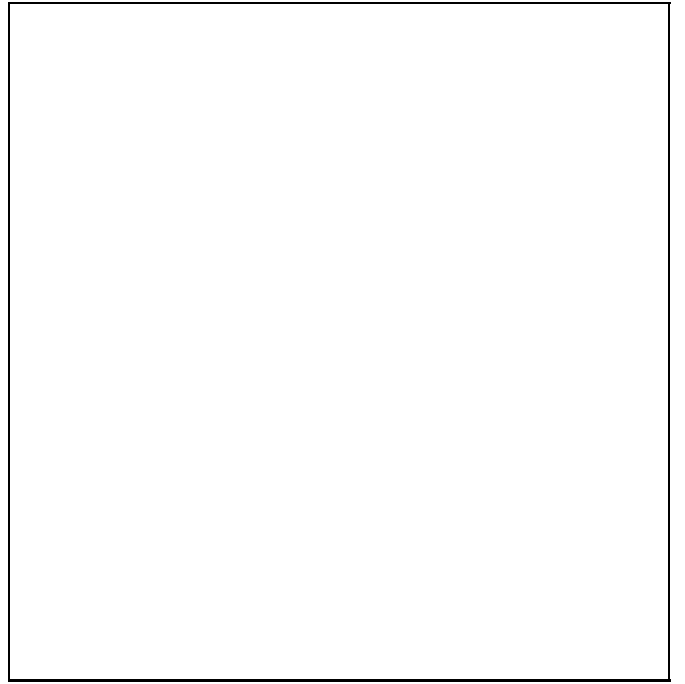

**Fig. 6.** Differential mass distribution $\mathrm{d}M/\mathrm{d}\Sigma$ and cumulated mass $M$ as a function of $\log\Sigma$, in density A with $L = 1$, $N = \infty$ (classical smooth but singular cloud model, Eq. (36))

The computed cloud being free of optical depth effects, the cloud can be characterised by a distribution of pixel averaged surface densities $\Sigma$. By comparing the different surface density distributions for different fractal dimensions, we can estimate in which situation the presence of a large range of surface densities, and large surface densities ($\sim$ large optical depths) in a small fraction of the surface, are likely to induce systematic errors on the mass estimate.

The surface density $\Sigma$ is calculated by counting the number of cloudlets in each pixel. In order to compute the differential mass distribution $\mathrm{d}M/\mathrm{d}\Sigma$ as a function of $\Sigma$, we have

$$\frac{\mathrm{d}M}{\mathrm{d}\Sigma} = \frac{\mathrm{d}M}{\mathrm{d}S}\frac{\mathrm{d}S}{\mathrm{d}\Sigma} = \frac{\mathrm{d}S}{\mathrm{d}\ln\Sigma} \ , \tag{41}$$

where $\mathrm{d}S$ is the surface element. So $\mathrm{d}M/\mathrm{d}\Sigma$ can be obtained by building the histogram of the number of pixels $\mathrm{d}S$ in intervals $\mathrm{d}\ln\Sigma$. The cumulated mass $M$ at lower $\Sigma$ is then obtained by integration: $M(\Sigma) = \int_0^\Sigma (\mathrm{d}M/\mathrm{d}\Sigma')\, \mathrm{d}\Sigma'$.

Similarly the differential area distribution $\mathrm{d}S/\mathrm{d}\Sigma$ as a function of $\Sigma$ is obtained by building the histogram of the number of pixels $\mathrm{d}S$ in intervals $\mathrm{d}\Sigma$, and the area covering factor of the mass at higher $\Sigma$ is obtained by integration: $S(\Sigma) = \int_\Sigma^\infty (\mathrm{d}S/\mathrm{d}\Sigma')\, \mathrm{d}\Sigma'$.

## 5. Results

Hundreds of different fractal clouds have been generated with $5 \leq N \leq 50$, and the number of cloudlets $N^L \lesssim 10^9$. Other variations on density laws not described here, or with varying

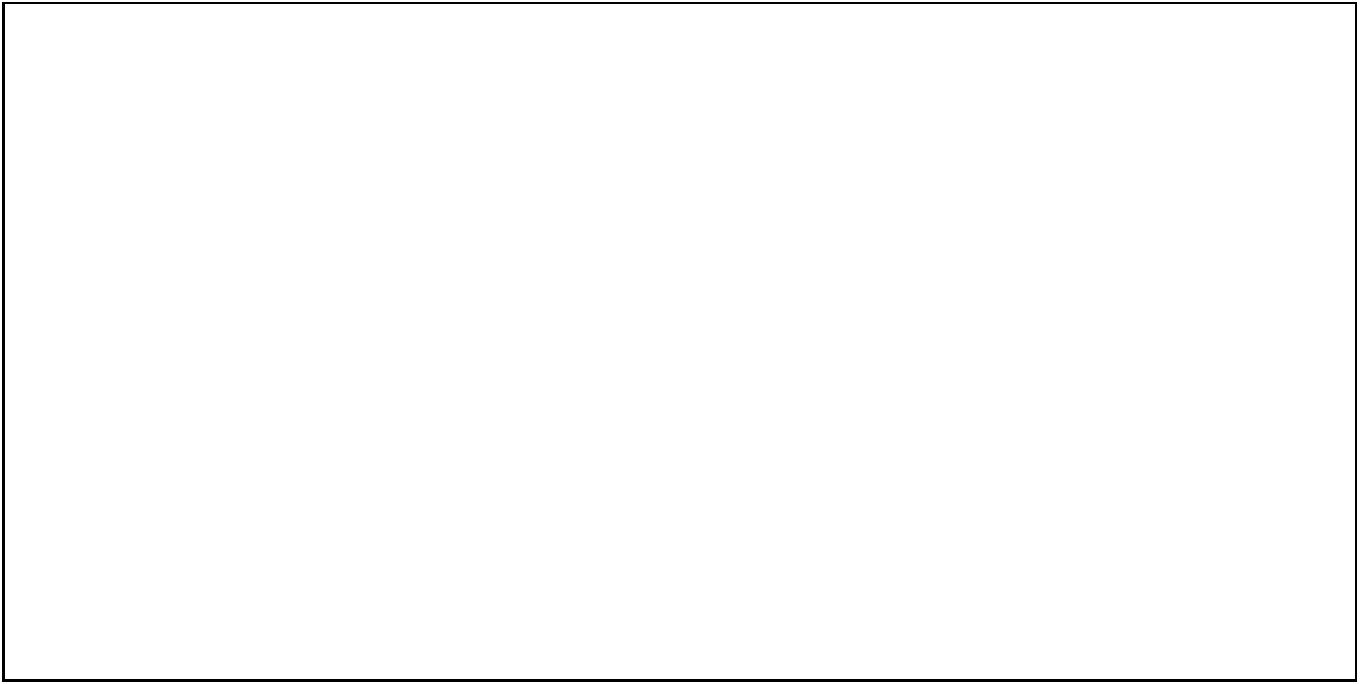

**Fig. 7.** Differential mass distributions $\mathrm{d}M/\mathrm{d}\Sigma$ as a function of $\Sigma$, for $D = 3$ (solid), 2.5 (dot), 2 (short dash), and 1.5 (long dash) and for the density laws A and B. Clearly lower $D$ models have more mass at higher $\Sigma$'s. The total mass is 1, and the total grid surface 4, which defines the unit of $\Sigma$

$N$, or with $N$ determined by a Poisson distribution at each level, have been experimented. The global results have been found to depend weakly on $N$, but strongly on $D$ and also on $\rho$. Below we describe only the models with $N = 10$ for density laws A and B.

### 5.1. Images

A few typical projected cloud models are shown in Fig. 5. The general aspect is indeed "cloudy". With $D = 3$ the area filling factor of a single cloud is large. Models resemble terrestrial clouds for $2 \lesssim D \leq 3$ because their area covering factor is near unity within well defined boundaries. With $D \approx 2$ the clouds have an intermediate area filling factor. When $D \lesssim 1.5$, the area filling factor drops rapidly, and one would need a large number of such clouds along the line of sight to cover a sizable fraction of the sky. This situation is precisely what is required in the Galaxy HI disc in order to have a small covering factor at high latitudes, but simultaneously at low latitudes 1) to be able to see far through the disc, 2) to have a near unity covering factor, and 3) to have optical depths around $\tau = 1$ or more (Burton 1992).

### 5.2. Surface density distributions

The distribution of mass as a function of surface density at a fixed resolution is mainly dependent on the dimension $D$. For example the mass distribution $\mathrm{d}M/\mathrm{d}\Sigma$ as a function of $\Sigma$ for the smooth truncated sphere with density A ($L = 1$, $N = \infty$) can be expressed in elementary terms as an implicit function of $r$ (cf. Eqs. (36), (41)),

$$\Sigma(r) = \frac{1}{2\pi r_L^2} \frac{\arccos\left(r/r_L\right)}{\left(r/r_L\right)} ,$$

$$\frac{\mathrm{d}M}{\mathrm{d}\Sigma}(r) = \frac{2\pi \left(r/r_L\right)^2}{\left[1 + \frac{(r/r_L)}{\arccos(r/r_L)\sqrt{1-(r/r_L)^2}}\right]} , \qquad (42)$$

and is shown in Fig. 6. Most of the mass is contained at low surface density.

In comparison, the mass distribution $\mathrm{d}M/\mathrm{d}\Sigma$ of fractal models are shown in Fig. 7 for different $D$'s, $N = 10$. Clearly for a fixed resolution more and more mass is found at higher $\Sigma$ as $D$ decreases. *The amount of mass at high surface density increases rapidly as $D$ decreases.* To first approximation these distributions resemble $\Sigma^a \exp[-(\Sigma/\Sigma_0)^b]$ distributions, where the constants $a$, $b$ and $\Sigma_0$ are in the range $0.5 \lesssim a \lesssim 2$, $1 \lesssim b \lesssim 2$ and $1 \lesssim \Sigma_0 \lesssim 10$.

In Fig. 8 we show the same differential mass distributions in $\log \Sigma$, enhancing low $\Sigma$'s. Visible spikes are due either to noise at the lowest $\Sigma$ and the discrete number of levels at higher $\Sigma$. Also shown is the cumulated mass $M$ at lower $\Sigma$ and the covered area $S$ at higher $\Sigma$ as a function of $\log \Sigma$. As typical for many unimodal distributions, most of the mass ($M > 0.5$) is reached at densities *larger* than the distribution maximum. Half of the mass is reached at surface densities about $3 - 7$ times higher with $D = 2$ than with $D = 3$ and about the square of that with $D = 1.5$. Also shown in Fig. 8 is the area covering factor

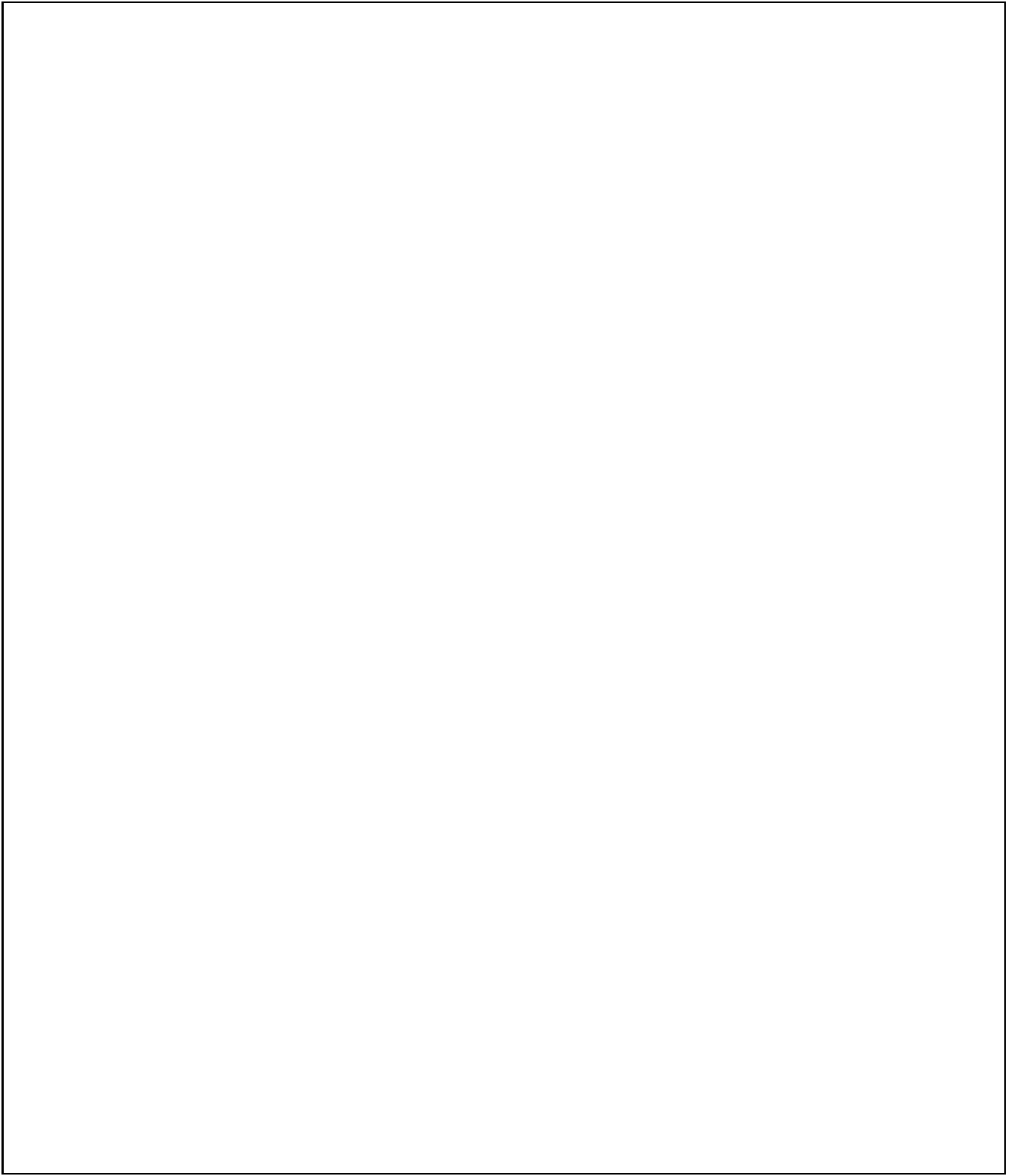

**Fig. 8.** Differential mass distributions $\mathrm{d}M/\mathrm{d}\Sigma$, cumulated mass $M$ at lower $\Sigma$, and area filling factor $S$ (in log) at higher $\Sigma$, as a function of $\log\Sigma$, for the same models as in Fig. 7

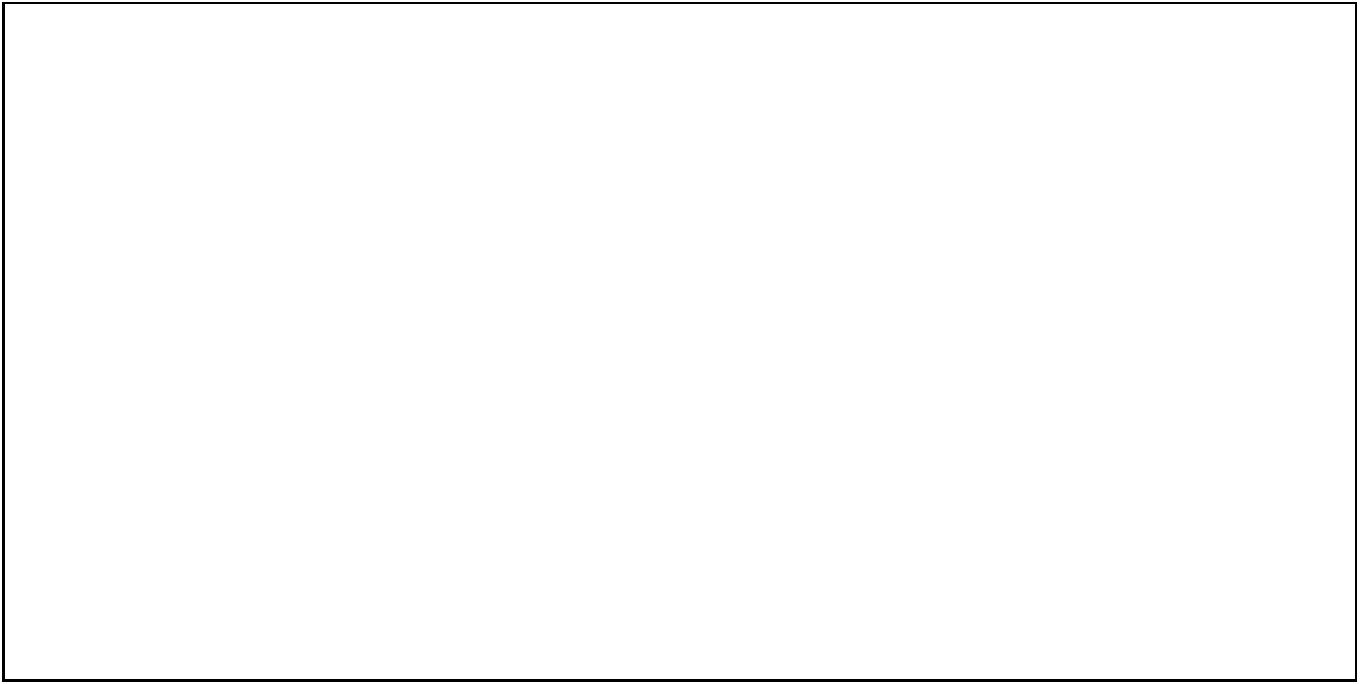

**Fig. 9.** Ratios (in log) of the $D = 3$ to the $D = 2.5$ (dot), $D = 2$ (dash) and to the $D = 1.5$ (long dash) cumulated mass in density laws A and B as a function of $\log \Sigma$

of higher $\Sigma$'s. Clearly as $D$ decreases half of the mass is rapidly contained in a very small fraction of the area: for density A at $D = 3, 2, 1.5$ the half mass surface is 0.257, 0.067, and 0.009 respectively, and for density B at the same $D$'s, 0.127, 0.019, and 0.003 (the total grid surface is 4, and the total mass is 1).

$M(\Sigma)$ can be considered as the mass that would be measured at lower $\Sigma$ because the medium would be optically thick at higher $\Sigma$ to a given wavelength. The ratio $M_3/M_D$ of the cumulated mass with $D = 3$ and lower $D$ is an indicator of the systematic error than would be made if one measures mass up to a finite $\Sigma$ due an optical depth limitation, and one thinks to be able to correct for optical depth effect and finite resolution because one assumes that the gas cloud has the structure of a "standard cloud" with $D = 3$. The ratios $M_3/M_D$ are relatively *slowly varying* at sufficiently low $\Sigma$'s (Fig. 9). For $D = 2$ we get $M_3/M_D \sim 4 - 16$, and much more at lower $D$ ($M_3/M_D > 200$ for $D = 1.5$). This means that if mass at higher density is not observed due to e.g. too large optical depths, we underestimate the mass by a factor $4 - 16$ if $D = 2$ when assuming $D = 3$, and even more if $D < 2$.

We would be able to measure all the mass independently of $D$ only in the case that the medium would be optically thin throughout the range of surface densities. Clearly observations at a single wavelength such as the 21 cm line are likely to miss much mass if the range of $\Sigma$'s spans several decades. In contrast, molecular clouds are observed at several wavelengths, and for them the virial mass is assumed to be $H_2$, so molecular clouds have by convention no missing mass problem.

### 5.3. Resolution effect

Since our models are far from covering a range of scales as large as the cold ISM is, it is important to examine how the mass distributions vary with the resolution. When the number of pixels is decreased, the mass distribution $\mathrm{d}M/\mathrm{d}\Sigma$ of a given model is found nearly invariant with $D = 3$, and tending toward the $D = 3$ distribution for lower dimensional clouds (Fig. 10). This means that an insufficient resolution is not important when $D = 3$, but becomes increasingly crucial when $D < 3$. In other words, a limited resolution is showing a restricted range of surface densities, mimicking a $D = 3$ distribution.

Since our higher resolution (2048) is only about the square root of the likely scale range in the cold ISM ($100\,\mathrm{pc}/20\,\mathrm{AU} \approx 10^6$), in order to evaluate the mass distributions over this range we can assume that as much effect would be observed between a resolution of $10^6$ and $10^3$ than between 1024 and $32 = \sqrt{1024}$, as in Fig. 10. If this is the case, then the estimates made from Fig. 8 should be corrected as follow: in a fractal distribution with a range of scales of $10^6$, half of the mass would be reached for surface densities about 20 times higher with $D = 2$ than with $D = 3$, and 400 with $D = 1.5$.

### 5.4. Density effect

The halo outside the characteristic radius $r_L$ in density B somewhat mimics an interclump medium. Its effect is to increase the width of the $\mathrm{d}M/\mathrm{d}\Sigma$ distribution by a factor $\approx 7$, in which more mass is shifted at high than at low density. So for $D = 1.5$ and $D = 2$ the half mass ($M = 0.5$) is reached at a surface density $\sim 2.6$ times *higher* in case B than in case A.

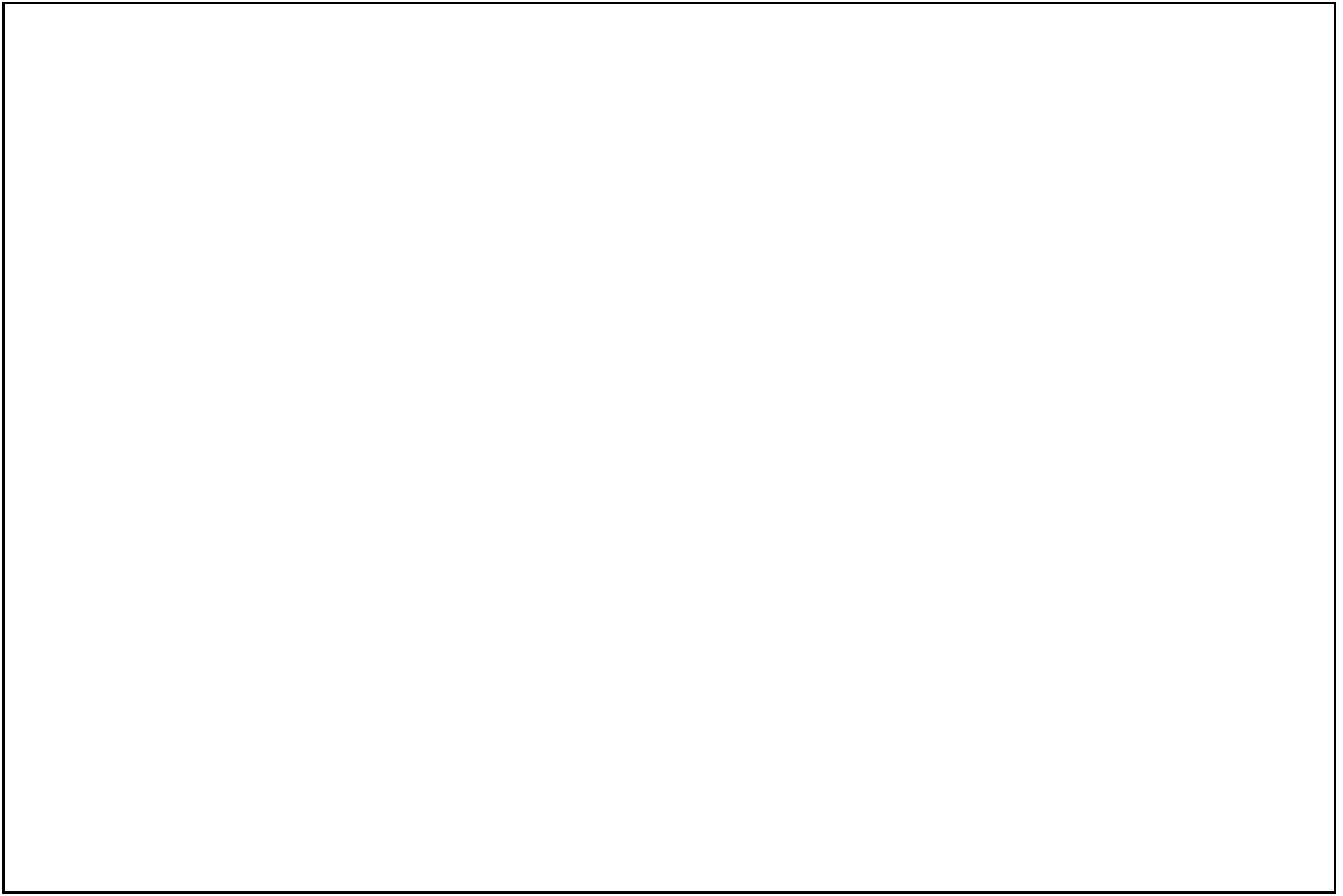

**Fig. 10.** Resolution effect in density law A, $D = 1.5$, 2, and 3, when increasing the number of pixels from $32^2$ to $1024^2$ for the same cloud model and shown for $\mathrm{d}M/\mathrm{d}\Sigma$ and $M$. An increasingly large amount of mass is shifted toward higher $\Sigma$'s at $D < 3$, while the mass distribution is practically *invariant* when $D = 3$

### 5.5. Summary

Geometric projection effects due to a fractal dimension lower than 3 are found to lead to a possible significant underestimate of the mass by single line observations. This error would be first due to the limited range of column densities over which a single line such as the 21 cm line is optically thin. Second high column density clumps covers a very small area of the sky, yet they can contain most of the mass. An underestimate by a factor 10 is typical in the models when $D \approx 2$, and rapidly more for $D < 2$. A factor 10 or more of hydrogen mass underestimate is enough to remove the need of exotic matter in disc galaxies (cf. Casertano & van Albada 1990).

A fractal cloud with $D < 3$ tends to *correlate the positions and velocities* of the individual cloudlets and simultaneously tends to have a low covering factor. When $D < 2$, the individual smallest fragments are more and more directly visible but the smallest fragments are increasingly denser in projection (cf. Eq. (11)). So at low $D$ most of the radiation coming from behind can peer through, while a large amount of mass may be hidden in dense clumps.

Physical effects may also induce systematic blindness to high surface densities. Since the average density of a clump of size $r$ scales as $\langle \rho \rangle \sim r^{D-3}$, when $D < 3$ the clump average density increases strongly at small scales. Then in addition to large optical depths, both the formation of $H_2$ and a low temperature close to 3 K may considerably increase the amount of mass not emitting the 21 cm line. The emission in any wavelength from the smallest clumps, the clumpuscules, would be virtually zero because they would be at a temperature of 3 K. The absorption by external radiation would be difficult to detect because the covering factor of a cloud by clumpuscules would be small, a few percents.

In our estimates no account has been taken of the clump velocity dispersion, that makes the medium more transparent. However, the opacity of the medium becomes so large for $D < 2$, that we expect a large residual opacity, even after taking velocity dispersion into account. Furthermore, we have seen that in a hierarchical fractal distribution the subclump velocities are strongly correlated to their parent clump velocity, and at small scale the velocity dispersion decreases, both effects conspiring to increase the optical depth. The fact is that both HI and CO observations are often made around $\tau = 1$, as conventionally estimated. Therefore, the velocity dispersion does not prevent these lines to be optically thick in observations. Clearly, more detailed models of the radiation transfer in a dynamical fractal medium would help in evaluating properly the effect of velocity dispersion.

## 6. Discussion

We discuss below a few problems and speculations connected with the hypothesis of dark matter being cold gas.

### 6.1. Possible dynamical and thermal equilibrium of the fractal

We have seen above (Eqs. (18), (21)) that in a hierarchical distribution with $D \gtrsim 1.5$ and small $N$ the clumps collide frequently at all levels. An actual smooth hydrodynamical behaviour happens only at the lowest level of fragmentation, where the gas becomes adiabatic.

Supersonic collisions of higher level clumps do not imply necessarily strong *gaseous* shocks, because higher level clump encounters are mainly gravitational. Before two clumps "collide", they first interact gravitationally. When their positions overlap in space, their subclumps can in turn "collide", repeating the process at a lower level. By the virial theorem, the relative initial velocity of two merging clumps acts to *cool* the final clump after a few $\tau_{\rm dyn}$, because for finding a new virial equilibrium the final clump needs to expand in average, since twice the initial kinetic energy excess is absorbed by gravitational energy.

The exact energetic budget following clumpuscules collisions is not obvious to determine, and would require specific simulations. When two clumpuscules collide, Eq. (21) shows that the collision is supersonic because $v_0$ is of the order of the clumpuscule sound velocity. For $2 < N < 20$ and $1 < D < 2$ the Mach number is always smaller than 2.12 so a typical shock is not hypersonic. Since the clumpuscules have $\tau_{\rm dyn,0} = \tau_{\rm ff} = \tau_{\rm KH}$, the heat propagates as fast as the clumpuscules expands. The momentary increase in temperature is not large because even in the conditions of a pure adiabatic shock (see e.g. Elmegreen 1992, Chap. 4.2) would lead to a temperature increase by a factor $< 3$ for $\mathcal{M} < 2$. In fact, the clumpuscule expansion, the near isothermality, and especially the virial theorem which requires that temperature *drops* after a few dynamical times in a gravitating system receiving energy, contribute all to decrease the amount of energy radiated away. Numerical simulations of supersonic cloud collisions have been shown to be surprisingly little dissipative due to the mainly gravitational character of the interaction (Lattanzio et al. 1985; Lattanzio & Henriksen 1988).

The important feature to retain from this is that clumpuscule collisions can both loose or gain energy from the bath. If the rate at which the energy radiated away by shocks is equal to the energy gained by the clumpuscules then the fractal gas can be in a statistical equilibrium for indefinite time, constantly fragmenting and forming clumpuscules.

### 6.2. Coupling with the 3 K radiation and solid $H_2$

A very important aspect of the physics of clumpuscules is the nature of their radiative coupling with the background radiation. Some relevant rotational lines from the gas are those of LiH and HD, at temperatures equal a few tens of K, since they are optically thick for the column densities of the clumpuscules (Lepp & Shull 1984).

But as soon as the background temperature falls below about $5 - 10$ K, the coupling may be ensured by $H_2$ ice, since a solid can radiate and absorb as a black-body. At these low temperatures, the $H_2$ molecule is mostly in the form of para-hydrogen. If the $H_2$ partial pressure exceeds the sublimation pressure, it may be in equilibrium with a kind of $H_2$ snow, provided ice crystals can start to grow on nucleation sites.

The possibility of the existence of solid $H_2$ in the ISM has been proposed by van de Hulst (1949) and Wickramasinghe & Reddish (1968), and a coupling of solid $H_2$ with the microwave background by Hoyle et al. (1968). Using better data on the sublimation curve this possibility was found unlikely by Greenberg & de Jong (1969) unless the hydrogen density exceed $10^5$ cm$^{-3}$, which holds in clumpuscules, but in which the amount of dust is unknown. The possibility of $H_2$ ice grains was also rejected by Field (1969) with the argument that the equilibrium temperature of approximately spherical grains with radiation is much higher than 3 K. But this argument fails if ice grains have a fractal (Wright 1993) or snowflake shapes.

More recently Sandford & Allamandola (1993) discussed from laboratory data the infrared properties of $H_2$ ices, also speculating that some of the "missing" mass might be solid $H_2$. They evaluate the critical temperature $T_{\rm crit}$ below which $H_2$ molecules can stick on an $H_2$ ice layer on the top of a ($H_2O$ rich) grain (their Eq. [5]). Replacing with the numerical values suited to this case, we find,

$$T_{\rm crit} \approx \frac{100}{56.46 - \ln n_{H_2}} \,. \tag{43}$$

For $n_{H_2} > 10^9$ cm$^{-3}$ we have thus $T_{\rm crit} \gtrsim 2.8$ K. The near coincidence with microwave background temperature is astonishing, because in the frame of the Big Bang theory it would not have occurred in the past.

From cryogenic industry tables (L'Air Liquide 1976) the sublimation curve of para-hydrogen can be approximated by

$$P_{\rm s} \approx 5.7 \cdot 10^{20}\, T^{5/2} \exp\left(-91.75/T\right) \; [{\rm K\,cm^{-3}}]\,, \tag{44}$$

for $T$ between 1 K and the triple point temperature at $T = 13.8$ K. This curve is very steep; it varies by more than 10 orders of magnitude between 2 and 4 K, and by about *40 orders of magnitude* between 1 and 13.8 K. In contrast Hegyi & Olive (1986) reject the possibility of solid $H_2$ because they assume, in particular, that the saturation pressure is *constant* and equals to the triple point pressure (70.4 mbar $= 5.1 \cdot 10^{20}$ K cm$^{-3}$), leading to the conclusion of rapid sublimation.

On the other hand the clumpuscule gas pressure $P_\bullet$ is taken from Eq. (3). We find that below about $3.1 - 3.3$ K, $P_\bullet$ is higher than $P_{\rm s}$. At $T = 3$ K, $P_\bullet/P_{\rm s} \approx 2 - 20$ and increases to about 17 times more at $T = 2.736$ K. The saturated pressure is reached in most of the clumpuscule. The question of the actual amount of solid $H_2$ depends critically on the presence of condensation sites such as dust, that we are presently unable to quantify for the outer part of galaxies.

Further complications can arise as in terrestrial clouds (see e.g. Pruppacher & Klett 1978): for example the progressive sublimation of falling snowballs due to the increasing temperature with depth, and the heat generated by friction of falling condensations depends strongly on the shape of the condensations (flakes, fractal grains, etc.).

Even if almost all the $H_2$ freezes out in some form of snow (some fraction of vapour $H_2$ remains as a partial pressure), the remaining He can sustain small $H_2$ grains by viscosity. But formally if the grain random velocity contributes to the pressure, the "molecular" weight $\mu$ increases to large values. In Eq. (3) the clumpuscule equilibrium mass $M_\bullet$ depends the fastest on $\mu$, and decreases to zero at large $\mu$, i.e., it dissolves in small, icy units resembling perhaps comets.

We estimate that for a population of snowballs of radius $r_*$ with the density $\rho_*$ of solid $H_2$ ($\approx 0.07\,\mathrm{g\,cm}^{-3}$), the photon mean-free path $\approx (n_* \pi r_*^2)^{-1}$ is smaller than the clumpuscule radius $R_\bullet$ when $r_* < (9/16)\beta\Sigma_\bullet/\rho_* \approx \beta \times 30$ cm, where $n_* = \beta\rho_\bullet/m_*$ is the number density of snowballs, and $\beta < 1$ is the fraction of the clumpuscule mass transformed into snow. So for typical snowball radii below, say, $1\,\mu$m, a clumpuscule is optically thick to optical or mm wave radiation even for $\beta$ as small as $10^{-3}$.

In any case in optically thick conditions the latent heat liberated by freezing ($\approx 100$ K/$H_2$) has to be transported out of the clumpuscule for further freezing, and this occurs not faster than $\tau_{\mathrm{KH}}$. Since the total latent heat is about 33 times larger than the thermal energy (3 K), necessarily the clumpuscule contraction is correspondingly slowed down, similarly to stars by nuclear reactions. Furthermore, if the clumpuscule effective temperature $T_{\mathrm{eff}}$ (i.e., the temperature at a depth where $\tau = 1$) is almost the background temperature, the outgoing flux is reduced by a factor $1 - (T_{\mathrm{back}}/T_{\mathrm{eff}})^4$ with respect to a heat loss without background radiation. Then the contraction time is much longer than the clumpuscule collision time $\tau_{\mathrm{col},0}$ since $\tau_{\mathrm{col},0}$ is of the order of $\tau_{\mathrm{KH}} = \tau_{\mathrm{dyn},0}$ (Eq.(19)); before a clumpuscule freezes out, a collision, expected to be slightly supersonic (Sect. 6.1), should reheat it by a few degrees during a time shorter than $\tau_{\mathrm{dyn},0}$, and the solid $H_2$ should sublimate.

We note that clumpuscules radiating at the effective temperature $T \lesssim 3.3$ K would be exceedingly difficult to observe against the microwave background at $T_{\mathrm{MWB}} = 2.736$ K, essentially because the fraction of the sky covered by the clumpuscules in galaxies at a redshift $z < 0.2$ (i.e. when $T_{\mathrm{MWB}} = 3.3$ K) covers much less than 1% of the galaxy surface when $D \approx 1.7$. The luminosity per unit mass, $4\sigma(T^4 - T_{\mathrm{MWB}}^4)/\Sigma_\bullet$ would be less than $0.004\,\mathrm{erg\,g}^{-1}\,\mathrm{s}^{-1}$. Since the mechanical energy lost by a self-gravitating system is just the actual kinetic energy, the time-scale to reach the present rotational kinetic energy with the above dissipation rate would be larger than $0.5\,v_c^2/0.004 \approx 2$ Gyr. This lower bound would be increased if we take into account the hotter $T_{\mathrm{MWB}}$ in the past, and possible energy inputs by residual cosmic rays or ionising radiation. So a clumpuscule luminosity of $0.004\,\mathrm{erg\,g}^{-1}\,\mathrm{s}^{-1}$ in the mm band can be sustained for a few Gyr by galactic rotation energy alone cascading the fractal down to the smallest level.

### 6.3. Turbulence dissipation time and outer disc energetics

Following Fleck (1981), let us calculate the largest scale $r_L$ at which the general galactic differential rotation can transfer energy into the gas by turbulence. The specific power generated by differential rotation in a turbulent medium is (Landau & Lifshitz 1971)

$$p_{\mathrm{g}} \approx \nu_L \left( R\frac{d\Omega}{dR} \right)^2 , \qquad (45)$$

where $\nu_L \equiv v_L r_L$ is the turbulent kinematic viscosity coefficient at the level $L$, $R$ is the galactic radius, and $\Omega$ is the galactic angular velocity. For a constant rotation velocity $v_c = \Omega R$ we have

$$p_L \approx v_L r_L \left( \frac{v_c}{R} \right)^2 = v_L r_L \Omega^2 . \qquad (46)$$

Equating $p_{\mathrm{g}}$ to $p_L$, and using Eq. (35) with $p_0 = v_0^5/G$, we get

$$\frac{r_L}{r_0} = \left( \frac{v_0}{v_c}\frac{R}{r_0} \right)^{2/(3-D)} \quad \text{or} \quad D = 1 + 2\frac{\ln\left( \dfrac{v_c r_L}{v_0 R} \right)}{\left( \ln\dfrac{r_L}{r_0} \right)} . \qquad (47)$$

For $R = 15$ kpc, $r_L = 1.5$ kpc a typical large scale turbulence scale, $r_0$ given by clumpuscules values $R_\bullet$ (Eq. (4)), $v_0 = \sqrt{3kT/\mu m_{\mathrm{p}}} = 0.18\,\mathrm{km\,s}^{-1}$, and $v_c = 200\,\mathrm{km\,s}^{-1}$, we get $D = 1.60$.

The time scale $\tau_{\mathrm{diss}}$ to dissipate the specific energy given by the rotation velocity ($= \frac{1}{2}v_c^2$) is therefore

$$\tau_{\mathrm{diss}} = \frac{1}{2}\frac{v_c^2}{p_{\mathrm{g}}} = \frac{1}{2}\tau_{\mathrm{dyn},0} \left( \frac{R}{r_L} \right)^2 \left( \frac{r_L}{r_0} \right)^{(3-D)/2} . \qquad (48)$$

For the same values of the parameters as above, we get $\tau_{\mathrm{diss}} = 3.7$ Gyr.

Considering now the maximum dissipation rate at the level of clumpuscules, we can roughly estimate the shortest time needed to dissipate energy of the clumpuscules by only blackbody radiation in order to reach the present bound disc state. The energy per gram that had to be dissipated by cold gas starting as a weakly bound gas at rest is just the present kinetic energy per gram, $\sim \frac{1}{2}(200\,\mathrm{km\,s}^{-1})^2 = 2 \cdot 10^{14}\,\mathrm{erg\,g}^{-1}$, so much more than the energy released by $H_2$ formation ($2.2\,\mathrm{eV/H} \approx 2 \cdot 10^{12}\,\mathrm{erg\,g}^{-1}$). The typical time scale for thermal energy radiation is then, using the expressions for $M_\bullet$ and $L_\bullet$ in Eq. (3)

$$\tau_{200} = \frac{\frac{1}{2}(200\,\mathrm{km\,s}^{-1})^2}{L_\bullet/M_\bullet} = 16\,T^{-9/4}\mu^{1/4}f^{1/2} \quad [\mathrm{Gyr}] . \qquad (49)$$

For $T = 3$ K, $\mu = 2.3$, and $f = 1$ we get $\tau_{200} = 1.2$ Gyr.

Thus, when neglecting incoming radiation, within a couple of Gyr cold gas discs can condense and get their observed rotation velocities by black-body radiation. This is a lower bound since in practice the pervading background 3 K radiation and near isothermal conditions decrease dissipation, and further energy sources are possible as discussed below.

In this scenario, the energetics of a pure cold gas disc can be understood to first order by a transfer of the kinetic energy of large-scale motion gained by the general disc contraction along the fractal hierarchy downward to the clumpuscules and then radiated away.

The candidate radiations for energy input are essentially the UV and X-ray background, since the cosmic rays from the star-forming regions have disappeared at large distances from the center. A likely limit on the extragalactic UV flux has been estimated to be 700 times smaller than the microwave background (Wright 1992), but this flux could be 10 times higher. The X-ray background is about $2 \cdot 10^4$ times less than the microwave background (Fabian & Barcons 1992). In cgs units Wright's "nominal" UV flux is about $2 \cdot 10^{-5}$ erg cm$^{-2}$ s$^{-1}$. If the UV flux can reach the clumpuscules (which is more likely if $D < 2$), with $\Sigma_\bullet \approx 1$ g cm$^{-2}$, we have a specific heating of $2 \cdot 10^{-5}$ erg g$^{-1}$ s$^{-1}$, that could still be 10 times higher. The heating input by UV radiation seems to be a possible complementary source of heating, while the X-ray background has a negligible contribution.

An eventual source of energy is the ram pressure due to the motion of a galaxy through the intergalactic gas. The total power generated is $\approx \frac{1}{2} \rho_{IGM} v_{gal}^3 S_\perp$, where $\rho_{IGM}$ is the intergalactic gas density, $v_{gal}$ the relative velocity of the galaxy to the gas, and $S_\perp$ is the cross-section area of the galaxy perpendicular to the velocity vector. So the specific power is $\frac{1}{2} \rho_{IGM} v_{gal}^3 / \Sigma_\perp$ where $\Sigma_\perp$ is the surface density projected along the velocity vector. In the very optimistic view that all this power is redistributed inside the galaxy and not immediately radiated away, and the galaxy disc is perpendicular to the velocity vector, the average specific power for $\rho_{IGM} < 0.001$ H cm$^{-3}$, $v_{gal} < 1000$ km s$^{-1}$, and $\Sigma_\perp > 0.1$ g cm$^{-3}$ is less than $4 \cdot 10^{-3}$ erg s$^{-1}$ g$^{-1}$. Since this specific power is inversely proportional to the surface density, this energy goes principally towards the diffuse gas, which is then naturally kept hot. But if any substantial heating would be due to ram pressure, it would be asymmetric on the ahead side of the galaxy. Also this source of heating should strongly varies among galaxies depending on the angle between the velocity vector and the disc rotation axis, and especially on the environment. Ouside clusters $\rho_{IGM}$ is smaller by $2 - 3$ orders of magnitude and then the corresponding power too. So heating by ram pressure appears as a eventual heating source in comparison with the energy losses estimated in Sect. 6.2 only in particularly dense environments.

### 6.4. Conditions for star non-formation

Hence, a possible solution to the problem of explaining why cold gas survives in the outer discs for $\gg 10^8$ yr without forming stars while its cooling time at large-scale is $\ll 10^4$ yr is that the gas is close to being isothermal and in a statistical equilibrium between coalescence, fragmentation, and disruption along a hierarchy of clumps interacting mainly through gravitation. Galactic rotational energy cascades through turbulence into the fractal down to the lowest level, the clumpuscules, where the temperature is so close to the 3 K that the dissipation time may exceed several Gyr. Star or Jupiter formation is prevented at the lowest scale because collisions reheat the clumpuscules faster than they can collapse in the adiabatic regime.

If the ambient temperature rises the clumpuscules get more massive, smaller and have a shorter dynamical time (Fig. 1), which decrease the collision rate (Eq. (18)). For example, assume that the total mass of a cloud is $10^6$ M$_\odot$, its radius 30 pc, and that heating is fast enough to conserve these macroscopic quantities. The fractal dimension as a function of $T$ and $\mu$ is then, plunging in the clumpuscules mass and radius relations (3) in Eq. (6) (with $f = 1$)

$$D(T, \mu) = \frac{77.2 - \ln T + 9 \ln \mu}{42.5 + 3 \ln T + 5 \ln \mu}. \qquad (50)$$

Fig. 11 shows that $D$ falls regularly from $D = 1.67$ at $T = 3$ K and $\mu = 2.3$ down to $D = 1.08$ at $T = 3000$ K, and then drops below 1 at $T > 5000$ K and $\mu = 0.63$. The ratio of the clumpuscule collision time to their free-fall time $\tau_{col,1}/\tau_{dyn,0}$ (Eq. (20)) increases sharply, especially at large $N$ (cf. bottom frame of Fig. 11). A clumpuscule can spend more time in the adiabatic phase during which it gradually contracts, before a collision shakes and mixes it. Furthermore, at $D < 1$ the collisions are subsonic which is favourable for clumpuscule coalescence. Then denser structures such as stars can be formed. This scenario needs one principal factor to form stars: a temperature increase of the fractal distribution above $\sim 3000$ K. Various causes, such as nearby supernova explosions, or large scale galactic disturbances, can lead to a rapid increase of ambient temperature.

Beside temperature changes, metallicity effects following star formation may change considerably the conditions for solid $H_2$ formation, because dust grains offer the possibility to start $H_2$ freezing.

### 6.5. Optical to gaseous disc transition

Although a tight relation exists between HI and disc dark matter, between the centre and the edge of a galaxy the fraction of dark gas should increase with radius. The almost perfect isothermality is only approached in the outer parts of the galaxies, far from stellar heating.

A progressive evolution of the physical conditions in the gas medium, as temperature increases is then natural to consider. If the ambient temperature increases, star formation occurs, and the strong heating of the gas at a temperature much higher than the background breaks the isothermality condition and allows rapid cooling, and more dissipation at all scales.

There are two mechanisms that can make the outer gas flow towards the inner galaxy. First, over a time scale comparable

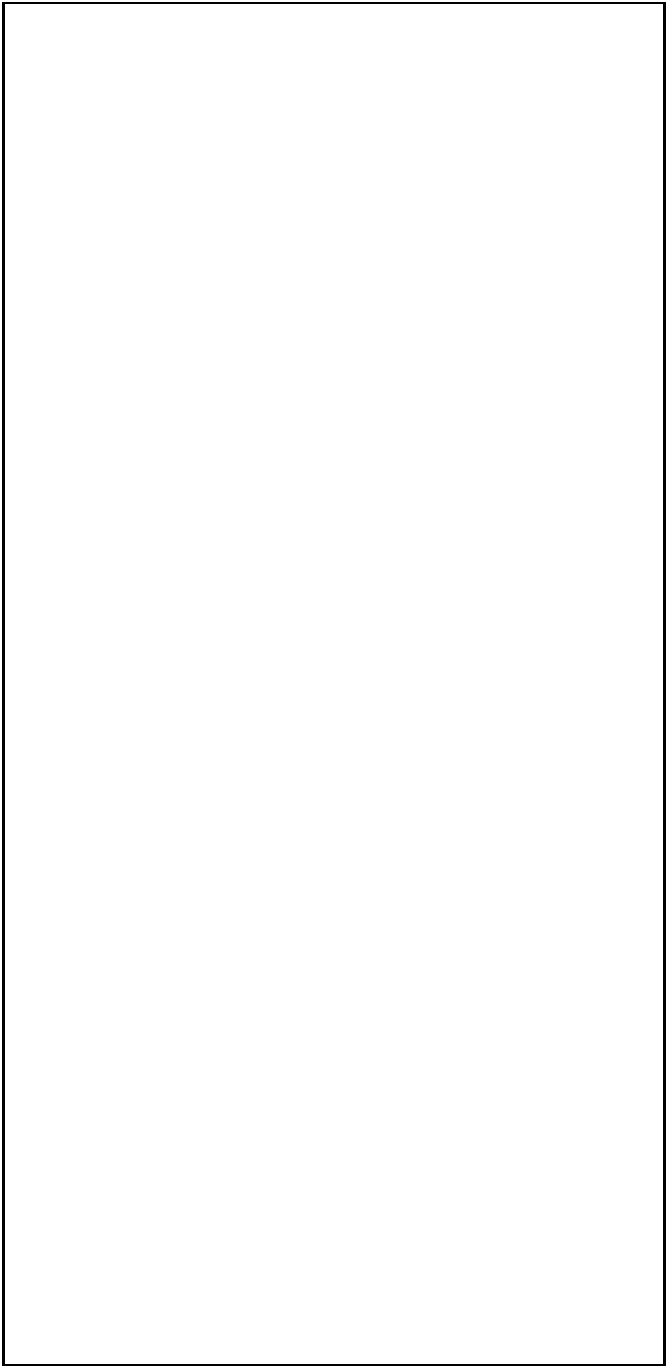

**Fig. 11.** Top: dependence of the fractal dimension $D$ as a function of $T$ for a cluster with $M = 10^6\,\mathrm{M_\odot}$ and a radius of 30 pc. At $T < 3000\,\mathrm{K}$, $\mu = 2.3$, and at $T > 5000\,\mathrm{K}$, $\mu = 0.63$. Bottom: dependence of the ratio $\tau_{\mathrm{col.1}}/\tau_{\mathrm{dyn.0}}$ as a function of $T$ for the same conditions, for $N = 10$ (solid) and $N = 100$ (dash)

with a Hubble time, the viscous torques slowly transfer the angular momentum of the gas outwards. Gravitational encounters of giant molecular clouds are able to increase the velocity dispersion at the expense of the rotational energy, and therefore to increase viscosity (Gammie et al. 1991). Second, more violent perturbations are provided by tidal interactions with compan-ions. At these occasions, strong gravity torques can drive the gas inwards in a dynamical time. This helps to understand the huge starbursts occurring in ultraluminous interacting or merging galaxies.

Increasing central densities in self-gravitating systems also means that a fraction of each clump evaporates. Thus at higher ambient temperatures, increasingly dense clumps are accompanied by an increasingly massive diffuse phase, with an increasingly large volume filling-factor, corresponding to the observations of nearby clouds. The no longer isothermal medium is then characterised by temperature and density gradients, several coexisting phases, and may contain more mass in a diffuse form. The physics of the interstellar medium inside the optical disc is then much more complex than the simple isothermal fractal picture in the outer parts. The neutral diffuse phase is visible in the HI 21 cm line, and metal-enriched denser regions are traced by molecular lines.

### 6.6. Fiedler clumps

The clumpuscules discussed here might be compared with the compact structures causing the extreme scattering events detected by Fiedler et al. (1987) at wavelengths of 3.7 and 11 cm. These extreme scattering events, lasting a couple of weeks, result presumably from the occultation of monitored quasars by Galactic AU-sized clouds not too far from the solar orbit. The scattering amplitude yields the electron density, found to be of the order of $4000\,\mathrm{cm^{-3}}$. Assuming a totally ionised medium Fiedler et al. deduce a very small mass of these objects ($\sim 10^{20}\,\mathrm{g}$).

Then a simple estimate shows that such a small completely ionised mass of this size at a temperature of $3\,\mathrm{K}$ should immediately explode, since its virial ratio is $\gg 10^7$. Now, if the detected electrons belong to a *partially ionised* dense and cold gas (the ionisation being induced by e.g. solar neighbourhood cosmic rays) the total mass can be grossly underestimated. Since Fiedler et al. mention that these clumps have sub-AU structures, the observed number of electrons has probably a complex relation with the effective mass. In dense clouds ($n \approx 10^3 - 10^7\,\mathrm{cm^{-3}}$) ionised by a constant rate of cosmic rays the ion density $n_{\mathrm{i}}$ is evaluated theoretically to be about $10^{-5}\,n_{\mathrm{n}}^{1/2}\,\mathrm{cm^{-3}}$, where $n_{\mathrm{n}}$ is the neutral density (Oppenheimer & Dalgarno 1974; Elmegreen 1979; Langer 1985). So the average value in an isothermal distribution ($\sim r^{-2}$) depends strongly on the low density parts which contain most of the mass; the result is sensitive to the truncation of the $r^{-2}$ distribution. Assuming instead a standard chemical composition and a virial gravitational equilibrium Eq. (3) yields for the "standard" 7 AU radius of Fiedler et al. a clumpuscule with a mass of $10^{-3}\,\mathrm{M_\odot}$ at a temperature of $15\,\mathrm{K}$.

The number density of these objects in the solar neighbourhood (independent of the virial equilibrium assumption) is estimated by Fiedler et al. by the time fraction of quasar occultation (0.7%), also the covered sky fraction. The estimated number density reaches the impressive value of $10^3$ times the star density. If their mass would be rather $10^{-3}\,\mathrm{M_\odot}$ each, then in the solar neighbourhood as much mass would be contained in these cloudlets than in stars! These cloudlets could therefore easily form all the controversial local dark matter advocated by Bahcall et al. (1992). The number density of cloudlets found by Fiedler et al. concerns the solar neighbourhood and is very approximate, because their distance could only be inferred to be Galactic.

### 6.7. Dark mass in cirrus clouds

We dont claim here that the mass of dense molecular clouds have been underestimated, since the $CO/H_2$ ratio has been empirically determined and justified from the virial assumption, but that the mass of HI clouds may be highly underestimated, based on the optically thin hypothesis for the HI emission. Because of confusion in the Galactic plane, in general HI clouds cannot be isolated; but at high latitudes, the situation is clearer. A few hundred clouds have been observed through HI emission (Heiles & Habing 1974), IRAS cirrus emission (Désert et al. 1988), or CO emission (Magnani et al. 1985) at high latitude, corresponding to clouds in the solar vicinity (at a distance of $\sim 100$ pc). From the CO size of these high latitude clouds (HLC) and the observed velocity dispersion, Magnani et al. (1985) concluded that these clouds are far from being gravitationally bound. Their stability would require that their real mass is $20 - 50$ times their observed gas mass. Since their dissolution time is of the order of $10^6$ yr, they concluded that they are extraordinarily young. Another interpretation of the data would be that the clouds are nearly virialised, but that we are underestimating the gaseous mass.

In Fig. 12 we draw the size-line-width relation for the high-latitude clouds (HLC) observed in CO by Magnani et al. (1985). We have computed the area of the clouds on their published maps, to derive the size in the same manner as Solomon et al. (1987). It can be seen that the observed line-widths and sizes of the HLC are consistent with an extension of the well-known $\sigma(v)$-size relation for dense virialised molecular clouds towards the small sizes. There is a lot more scatter for the HLC, probably because of more uncertainty in their distance. To draw the figure, we have assumed the same distance for all clouds, equal to the average value of 105 pc found by Magnani et al. (1985). It is then tempting to assume that HLCs are also approximately in virial equilibrium, and that their mass has been grossly underestimated. Interestingly, Penprase (1993) finds that these clouds reveal a distinct enrichment of molecular content with respect to their extinction $E(B - V)$, and that this could come from our ability to better resolve the densest structures in these very nearby objects.

### 6.8. Dark matter lensing

Can the clumpuscules provide microlensing effects, and be detected by the current search of "Massive Compact Halo Objects" towards the Large Magellanic Cloud (e.g. Moniez 1990)? According to Paczynski (1986), it is impossible to see significant amplification effects by a compact deflector for an impact parameter of the light ray larger than the Einstein radius $R_E = 2\sqrt{GMd}/c$; here $M$ is the mass of the deflector, $d = \overline{OD} \cdot \overline{DS}/\overline{OS}$, where O is the observer, D the deflector, and S the source. An extended deflector, of size larger than the Einstein radius, does not provide any lensing effects, but only obscuration effects. For stars in the LMC as the sources S, and a deflecting clumpuscule D at 10 kpc from the Sun, $R_E = 0.3$ AU, therefore no microlensing effect is expected. $R_E$ is even smaller by an order of magnitude for deflectors within 800 pc from the Sun, or within 800 pc from the source S. Therefore the eventual too rare detection of micro-lensing events will not prove the absence of baryonic dark matter in our neighbourhood.

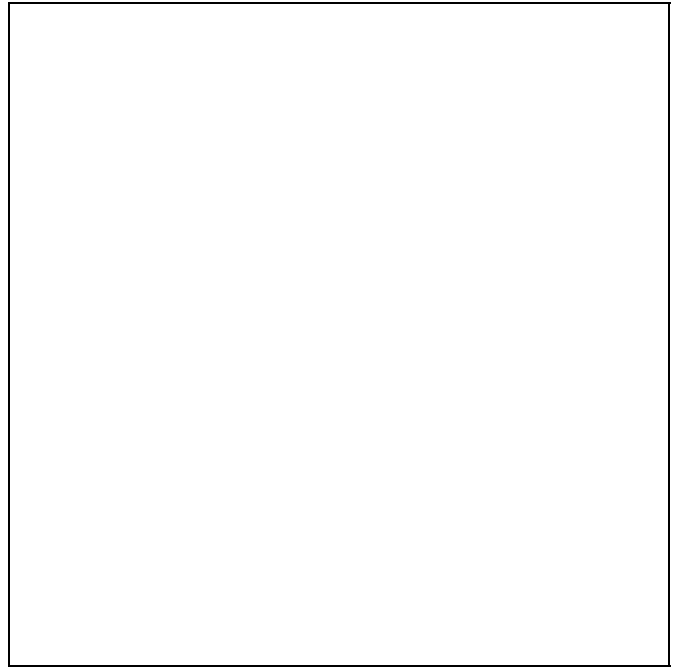

**Fig. 12.** Size-line-width relation for the high-latitude clouds (HLC, filled squares) observed in CO by Magnani et al. (1985), compared to the same relation found for galactic plane clouds by Solomon et al. (1987) (open squares, and solid line)

If the clumpuscules do not provide any microlensing effect, they might however be detected by their eventual obscuration effects if they contain some dust or ice grains. A source star in the LMC could disappear and reappear through the proper motion of the clumpuscule (at $V \approx 200$ km s$^{-1}$), in a time-scale of the order of a few months. The probability is low of course, but the order of magnitude can be derived from the Fiedler et al. (1987) scattering events. A rough estimate gives a probability of the order of $10^{-6}$ for a given source to be obscured. With a number of sources of the order of $10^6$, the experiment is within reach. For this obscuration experiment, it would be preferable to choose a region of the Magellanic Clouds already known to be partially covered by some Galactic cold interstellar matter.

### 6.9. Gas heating and cooling in clusters

In galaxy clusters, the frequent interaction between the gaseous halos progressively heats and destroys the fractal structure of cold gas. The isothermality is broken because the long cooling time ($\sim 1$ Gyr) of the hot phase with respect to the cold phase favours thermal instabilities. The mass of the hot coronal gas in clusters has been estimated from 2 to 8 times that of the visible mass (David et al. 1990; Edge & Stewart 1991), depending on the Hubble constant value as $H^{-3/2}$, and from cluster to cluster. The mass ratio between the hot gas and the stars increases with the temperature of the gas. Our hypothesis provides an easy interpretation for these observations: it explains in particular why the gas mass dominates the stellar mass, and predicts a more massive hot diffuse phase with a higher temperature. Indeed, as the interaction frequency between cold halos increases, temperature increases both from galaxy interaction and from the resulting increased star formation, and then the cold clumpuscules are more easily evaporated in the intercluster medium.

Clusters are therefore the privileged regions where dark matter is liberated from its cold dark phase, and where it becomes visible. But this diffuse phase is only transient. When the density of hot gas is high enough, the hot medium becomes unstable with respect to cooling, and a cooling flow starts up in the cluster core. Once cold, the gas would immediately fragment again in clumpuscules, and would disappear from view. This could solve the long known problem of disappearance of mass in cooling flows (Fabian 1987). Several solutions to the problem have been proposed; one is to assume that the cooling gas, due to the high ambient pressure, forms only low-mass stars or even planetary objects (Fabian et al. 1982), that are difficult to observe. Another solution was to try to build models of cooling flow with constant mass flux, that do not have to form stars. But these models did not succeed to be realistic and fit the observations (White & Sarazin 1988). They conclude that the mass flux must be decreasing inwards, that is, matter is constantly dropping out of the flow, and becomes dark. This scenario is supported by the recent discovery that dark matter in clusters is more concentrated that visible matter (Wu & Hammer 1993). The core radius of the matter is at least 10 times lower than that of the hot gas, which suggests a lot of dissipation.

When a cooling flow starts in the core of a cluster, thermal instability provokes the separation of at least two phases: a cold dense phase, and the hot medium in which it stays in pressure equilibrium (Fall & Rees 1985). Since the cooling rate is proportional to the square of the density, the cooling time becomes shorter and shorter for the dense colder phase, which is compressed to still higher densities by the hot medium. The latter stays at the virial temperature of the cluster. The cold phase experiences a fragmentation cascade, similar to that described in previous sections. The volume filling factor of this cold phase is negligible, and if the gas remains little enriched in metals, it is optically transparent. If the external pressure plays a rôle at the beginning of the condensation of the cold phase, it soon becomes negligible, and self-gravity takes over. The external pressure is of the order of $P = 4 \cdot 10^4 \, \mathrm{K \, cm}^{-3}$, for a hot medium of density $10^{-3} \, \mathrm{cm}^{-3}$, and temperature $4 \cdot 10^7 \, \mathrm{K}$, while the pressure inside a clumpuscule is of the order of $10^9 \, \mathrm{K \, cm}^{-3}$.

It is the hierarchical structure of the cold phase that prevents the clouds from evaporating. According to McKee & Cowie (1977), the evaporation time-scale of a clumpuscule isolated in a $T = 10^6 \, \mathrm{K}$ medium is of the order of $10^7 \, \mathrm{yr}$, so much longer than the collision time ($\approx 10^3 \, \mathrm{yr}$). At the interface, the X-ray photons from the hot gas penetrate over a mean free path, $(n_H \sigma_\nu)^{-1}$ of the order of $10^{10} \, \mathrm{cm} = 10^{-4} R_\bullet$, where $\sigma_\nu = 1.8 \cdot 10^{-20} (h\nu/150 \, \mathrm{eV})^{-3} \, \mathrm{cm}^2$ (Brown & Gould 1970). This leads to consider 3 phases in the intergalactic gas, beside the hot medium and the cold neutral phase, there is a warm ionised interface, bearing some similarities with the ISM model by McKee & Ostriker (1977). In the latter case, the three-components medium is in phase equilibrium, regulated by supernovae explosions. In clusters with cooling flows, the cold phase is still condensing out of the hot phase.

### 6.10. Overcooling problem in the early Universe

A similar overcooling problem occurs in the early universe. It was recently discussed in the frame of CDM cosmologies by e.g. Blanchard et al. 1992, although the problem arises in a baryonic universe as well: according to the well-known gas cooling functions at a given temperature, density and metallicity (e.g. Dalgarno & McCray 1972), the large structures that collapse at large virial temperatures, larger than $10^4$ K, are violently unstable. The gas cooling is almost instantaneous. As soon as the gas temperature is lower than the virial temperature of the potential in which it settles, it contracts, becomes denser and denser, and the cooling is more efficient. It is the cooling runaway. Since it was believed until now that all the cooling gas was bound to turn into stars in a short time-scale, all the baryons would have to be seen as stars now. The total amount of stellar matter, i.e. galaxies, is then predicted more than 10 times higher than is observed! To solve this problem, reheating mechanisms have been invoked, and the existence of large amounts of hot gas in clusters provided some evidence that this reheating was efficient (Blanchard et al. 1992).

The overcooling is easily solved by releasing the assumption that the cooling gas must turn into stars. As soon as the gas contracts, the hierarchical fragmentation proceeds, building a nearly isothermal fractal structure. Only a tiny fraction of this gas turns into stars, at the density and temperature peaks that correspond to galaxies. The outer cold dark gas would remain quiescent and stable for still a few Gyr, slowing contracting, and revealed only by galaxy-galaxy interactions.

### 6.11. Primordial chemistry

In the early Universe at $z < 1000$ the primordial gas after recombination should also be Jeans unstable and should have a highly inhomogeneous fractal phase during which the rate of formation of $H_2$ molecules (knowing that the relevant differential equations of chemical reactions are stiff) could be much larger than the ones found in homogeneous chemical models (e.g. Puy et al. 1993, and references therein). Since the neutral-neutral $H_2$ formation time-scale, which is a conservative upper bound, is $3 \times 10^6/n_H$ Gyr (see e.g. Genzel 1992) it is clear that density inhomogeneities of the order of $10^6 - 10^{10}$ modify completely the conditions at which chemical reactions take place.

# 7. Conclusions

In Paper I we have recalled that the arguments for the presence of dark matter in spirals have become weaker for the optical parts of galaxies, while a form of dark matter able to form stars is needed in outer gaseous discs.

Better observations of the ISM show that the cold gas is fractal and essentially clumpy down to very small scales, of the order of a few AU. A wide range of gas densities ($\gtrsim 10^7\,H\,cm^{-3}$) that continuity between dense clumps and stars leads to infer the existence, cannot be investigated by today's techniques. Our fractal models suggest that HI mass determination could underestimate the gas mass by a factor 10 or more owing to the very inhomogeneous nature of cold gas leading to high optical depths in a small fraction of the sky, rapid $H_2$ formation, and near thermal equilibrium with the 3 K radiation. If such a large error can be confirmed by independent mass determinations, then the problem of dark matter in discs would be solved. The dark matter to HI constant ratio would seem then more natural.

Physical considerations about the smallest fragmentation clumps in a fractal gas suggest that most of the mass in galaxies would then be in the form of cold ($\sim 3$ K) H+He clumpuscules, $\sim 30$ AU in size, with a mass of the order of a Jupiter, so much less dense than Jupiter. The physical state of these smallest structures of the fractal gas has been described. It turns out that the clumpuscules can both hide most of the mass (their average column density is $10^{24}\,H\,cm^{-2}$) and allow to explain why cold gas does not collapse immediately into stars or Jupiters.

An undetermined but probably small fraction of the mass might be in solid $H_2$ form. Indeed the clumpuscules have favourable conditions for $H_2$ freezing. But critical for freezing is the unknown amount of nucleation sites. The coincidence that the clumpuscule conditions are just favourable to build $H_2$ ice at a fraction of degree from today microwave background temperature is astonishing.

For an expected fractal dimension $D > 1.5$ and small $N \sim 10$ the clumpuscule collision rate is comparable to the clumpuscule contraction time in the adiabatic phase, preventing them to become much denser. The almost isothermal conditions decrease strongly energy dissipation to values low enough to be sustained by galactic rotation and extragalactic ionising radiation for several Gyr.

If galactic rotation energy is slowly transferred down to the fractal in a steady state, the specific power should be constant along the hierarchy. In these conditions the predicted fractal dimension of the fractal should be $D = 1.67$ (cf. Eq. (35)).

When temperature raises, the clumpuscules grow in mass and spend more time in the adiabatic collapse phase, while the collision time increases rapidly, and the collision speed becomes subsonic. Collisions become inefficient to stop collapse. A runaway process leading to star formation can follow.

*Acknowledgements.* We are grateful to Ron Allen, James Lequeux, and Louis Martinet for useful discussions. We thank John Scalo for refereeing thoroughly this work. In particular he pointed out the crucial aspect of solid $H_2$ phase. This work was partly supported by the Swiss National Science Foundation.